\documentclass[lettersize,journal]{IEEEtran}
\usepackage{amsmath,amsfonts}
\usepackage{algorithmic}
\usepackage{algorithm}
\usepackage{array}
\usepackage{booktabs}
\usepackage{caption}
\usepackage{svg}

\usepackage{textcomp}
\usepackage{stfloats}
\usepackage{url}
\usepackage{verbatim}
\usepackage{graphicx}
\usepackage{cite}
\usepackage{wasysym}
\usepackage{amssymb}
\usepackage{forest}
\usepackage{tikz}
\usepackage{tcolorbox}
\usepackage{xcolor}
\usepackage{hyperref}
\usepackage{subcaption}
\usepackage{amsthm}
\usepackage{makecell}
\usepackage{multirow}
\usepackage{threeparttable} % 如果需要表格注释
\usepackage{tikz}
\usetikzlibrary{positioning}

\newtheoremstyle{findingstyle} % Style name
  {3pt} % Space above
  {3pt} % Space below
  {\itshape} % Body font
  {} % Indent amount
  {\bfseries} % Theorem head font
  {.} % Punctuation after theorem head
  {.5em} % Space after theorem head
  {} % Theorem head spec
\theoremstyle{findingstyle}

\newtheoremstyle{summarystyle} % Style name
  {3pt} % Space above
  {3pt} % Space below
  {\itshape} % Body font
  {} % Indent amount
  {\bfseries} % Theorem head font
  {.} % Punctuation after theorem head
  {.5em} % Space after theorem head
  {} % Theorem head spec
\theoremstyle{summarystyle}

\newtheoremstyle{remarkstyle} % Style name
  {3pt} % Space above
  {3pt} % Space below
  {\itshape} % Body font
  {} % Indent amount
  {\bfseries} % Theorem head font
  {.} % Punctuation after theorem head
  {.5em} % Space after theorem head
  {} % Theorem head spec
\theoremstyle{remarkstyle}
\newtheorem{remark}{Remark}

\usepackage{ulem}

\begin{document}

\title{A Survey on Physical Adversarial Attacks against Face Recognition Systems}

%\title{Physical Adversarial Attacks against Face Recognition Systems: A Survey and Analysis}

\author{Mingsi Wang, Jiachen Zhou, Tianlin Li, Guozhu Meng, Kai Chen

\IEEEcompsocitemizethanks{
\IEEEcompsocthanksitem Mingsi Wang, Jiachen Zhou, Guozhu Meng, and Kai Chen are with the Institute of Information Engineering, Chinese Academy of Sciences, Beijing 100085, China, and also with the School of Cyber Security, University of Chinese Academy of Sciences, Beijing 100049, China. \protect E-mail: \{wangmingsi, zhoujiachen, mengguozhu, chenkai\}@iie.ac.cn. 
\IEEEcompsocthanksitem Tianlin Li is with Nanyang Technological University, Singapore. \protect E-mail: tianlin001@e.ntu.edu.sg.
}

}

\maketitle

\begin{abstract}
As Face Recognition (FR) technology becomes increasingly prevalent in finance, the military, public safety, and everyday life, security concerns have grown substantially. Physical adversarial attacks targeting FR systems in real-world settings have attracted considerable research interest due to their practicality and the severe threats they pose. However, a systematic overview focused on physical adversarial attacks against FR systems is still lacking, hindering an in-depth exploration of the challenges and future directions in this field. In this paper, we bridge this gap by comprehensively collecting and analyzing physical adversarial attack methods targeting FR systems. Specifically, we first investigate the key challenges of physical attacks on FR systems. We then categorize existing physical attacks into three categories based on the physical medium used and summarize how the research in each category has evolved to address these challenges. Furthermore, we review current defense strategies and discuss potential future research directions. Our goal is to provide a fresh, comprehensive, and deep understanding of physical adversarial attacks against FR systems, thereby inspiring relevant research in this area.
\end{abstract}

\begin{IEEEkeywords}
Face recognition, adversarial examples, physical world, attacks and defenses, AI security, deep learning.%, survey.
\end{IEEEkeywords}
\section{Introduction}
\label{section: introduction}
\IEEEPARstart{F}{ace} Recognition (FR) systems have made significant strides in both performance and scalability, driven by advanced deep learning techniques \cite{lecun2015deep, DeepFace, FaceNet}. 
These systems are now integral to various applications, including device unlocking\cite{patel2016secure}, e-banking \cite{wang2021exploring}, and military operations\cite{goel2019development}. FR systems primarily work by utilizing well-trained FR models to extract facial features from digital images or video frames captured by cameras, enabling the identification and authentication of individuals. However, security concerns have greatly intensified alongside the rapid development of FR technology. Recent studies \cite{szegedy2013intriguing, dong2019efficient, zhong2020towards} have revealed that FR systems are vulnerable to adversarial attacks, where adversarial samples are deliberately crafted to deceive the FR systems. %\ltl{Can find some news.}

Adversarial attacks against FR systems can be categorized into digital and physical attacks based on their domain of execution. Digital adversarial attacks occur after camera capturing and involve directly manipulating digital pixels, typically by adding imperceptible perturbations to face images to generate adversarial samples that deceive the FR model. Numerous studies\cite{dong2019efficient,zhong2020towards,yang2021attacks,deb2020advfaces,jia2022adv,goswami2018unravelling,zhong2019adversarial,chatzikyriakidis2019adversarial,dabouei2019fast} have proposed methods for generating subtle yet potent global perturbations across the entire face image. However, these global digital perturbations, while difficult to print and capture accurately, are impractical in real-world scenarios, where attackers are limited to modifying only the face without altering the surrounding background captured by the camera. This limitation has prompted researchers to explore more practical attack strategies in the physical world. 

In contrast to digital adversarial attacks which perturb pixels globally, physical adversarial attacks focus on objects around faces in the real world, manipulating them before the image is captured by the camera to deceive FR systems without directly interacting with the FR model. Various techniques have been proposed by attackers to introduce localized patches or patterns to the face, such as adversarial stickers\cite{FaceAdv, GenAP, RHDE, AdvSticker}, accessories\cite{AdvGlass, AdvMakeup, MAKEUP, AdvHat}, infrared lights\cite{IMA, invisible, LZP}, and illumination\cite{VLA, Relighting, LIM}. Despite these advancements, there remains a significant gap in providing a precisely focused overview that clearly clarifies the unique challenges of this field, comprehensively analyzes existing attack methods, and prospectively explores potential future directions. To address this gap, we present a systematic and in-depth survey of current physical adversarial attack methods targeting FR systems in this paper.

\subsection{Motivation}
The main motivations for this survey are as follows:

(i) FR, a contactless biometric technology widely applied in fields such as finance and public safety, is facing significant security challenges from adversarial attacks. Physical adversarial attacks, which have been proven effective in deceiving face recognition systems \cite{FSG}, are especially concerning in real-world environments, as they can compromise a wide range of already deployed security applications, posing significant risks to practical systems. Therefore, constructing a systematic survey of physical adversarial attacks against FR systems is highly valuable for both academics and industry. 

(ii) Physical adversarial attacks present unique challenges in real-world applications that distinguish them from digital adversarial attacks. Firstly, physical adversarial perturbations require a common real-world medium to function and should appear natural within the surrounding environment when applied. Secondly, ensuring effectiveness through virtual-to-real and real-to-virtual transformations is challenging. Thirdly, the difficulty and cost of implementing such attacks in real-world settings hinder their practicality. Fourthly, optimizing universal perturbations, which attain strong reusability across diverse input samples, remains a significant challenge. Finally, it remains challenging for physical adversarial attacks to achieve strong transferability across various custom-designed, commercial, and confidential FR systems deployed in real-world scenarios with black-box access. 
Thus, it is crucial to develop a thorough understanding of the challenges posed by the intrinsic attributes of physical adversarial attacks and effective solutions to address them.

(iii) While numerous studies on physical adversarial attacks against FR systems have been proposed, a systematic overview of existing physical attacks is still needed to summarize current advancements and chart future research directions. 
Although several related surveys have been published, they do not specifically focus on face recognition systems. 
Consequently, these surveys lack up-to-date investigations, comprehensive analyses, and in-depth insights specifically examining physical adversarial attacks against FR systems. For instance, surveys like \cite{wang2022survey, wei2022visually, guesmi2023physical, wang2023adversarial, wei2024physical, fang2024state, kong2022digital} explore physical adversarial attacks within the broader field of computer vision. The survey \cite{Surveillance} examines physical adversarial attacks in the context of surveillance systems. The study\cite{hasan2023presentation} investigates the application of physical adversarial attacks in face privacy protection. 
The work\cite{vakhshiteh2021adversarial}, while focused on FR systems, provides limited coverage of physical adversarial attacks and defenses but primarily emphasizes adversarial attacks and defenses in the digital domain. Additionally, while introducing attack and defense methods, this study focuses mainly on listing and categorization rather than a detailed analysis and comparison. Therefore, a more comprehensive and in-depth examination of physical adversarial attacks focusing exclusively on FR systems is needed.

\subsection{Research Questions}
This survey aims to provide an overview of the unique challenges in physical adversarial attacks against FR systems, the corresponding solutions proposed by existing attack methods, current defense mechanisms, and potential future research directions. Specifically, we answer the following questions:
\begin{itemize}
    \item \textbf{RQ1:} What are the unique challenges of physical adversarial attacks on FR systems? (answered in Section \ref{section: overview})
    \item \textbf{RQ2:} How do existing attack methods operate and address these inherent challenges? (answered in Section \ref{section: attack})
    \item \textbf{RQ3:} How are current defense mechanisms conducted? (answered in Section \ref{section: defense})
    \item \textbf{RQ4:} What are the potential future directions in this field? (answered in Section \ref{section: future directions})
\end{itemize}

\subsection{Collection Strategy} 

In this survey, we concentrate on physical adversarial attacks against FR systems. To gather a broad and representative range of studies, we employed a systematic search strategy using keywords such as ``physical attack'', ``physical adversarial examples'', or ``face recognition'' in Google Scholar. In addition, we manually reviewed the references of all selected papers to ensure comprehensive coverage of the relevant literature. 
Given the critical importance and high complexity of security issues in this field, related research remains relatively scarce. Consequently, we collect a comprehensive set of 40 relevant papers from authoritative conferences and journals spanning from 2016 to 2024, specifically focusing on physical adversarial attacks in the context of face recognition. 
Consequently, our study presents a more up-to-date, extensive, and in-depth analysis of the physical adversarial attacks against FR systems, offering crucial insights into this fast-evolving domain.

\noindent\textbf{Contributions.} Our contributions are summarized as follows:
\begin{itemize}
    \item We provide a comprehensive review of existing physical adversarial attacks against face recognition systems, highlighting their substantial threat in real-world scenarios.
    \item We explore the differences between physical and digital adversarial attacks based on their workflows and identify the unique challenges posed by physical attacks in terms of their key attributes.
    \item We categorize existing attack methods by their physical medium, analyze how they address the identified challenges, and summarize their strengths, limitations, and practical implications of each category.
    \item We conduct a thorough review of current defenses and discuss promising directions to facilitate future research.
\end{itemize}

The rest of the paper is structured as follows: Section \ref{Preliminaries} provides the background knowledge and benchmarks for physical adversarial attacks.
Section \ref{section: overview} outlines the key differences between physical and digital adversarial attacks against FR systems, and reveals the unique challenges posed by physical adversarial attacks. 
Section \ref{section: attack} categorizes existing physical adversarial attacks and summarizes their solutions to address the associated challenges.
Section \ref{section: defense} reviews current defenses against such attacks. Section \ref{section: future directions} discusses potential future research directions. Finally, Section \ref{section: conclusion} concludes the paper. 

\section{Preliminaries}
\label{Preliminaries}

\subsection{Notations and Terms}
\subsubsection{Notations}

In Table \ref{tab:Symbols}, we standardize and define the notations used throughout this paper. 

\begin{table}[htbp]
\centering
\caption{Notations and their definitions.}
\tabcolsep=0.05cm
\label{tab:Symbols}
% \scalebox{0.87}
{\begin{tabular}{cc|cc}
\toprule
\textbf{Notation}  & \textbf{Definition}                            & \textbf{Notation}               & \textbf{Definition}      \\ \hline
$f_\theta$        & FR model            &  
$\mathbb{E}_{t_1,t_2\thicksim T_1,T_2}$ & EoT transformation                \\
$x$               & Benign image                & 
$f_{d2p}$         & D2P module                  \\
$x^{t}$           & Target image                  & $\mathcal{L}(x,p)$                      & Total loss                        \\
$x^{adv}$         & Adversarial example           & $\mathcal{L}_{adv}$                     & Classifier loss                   \\
$\mathcal{M}$     & Binary mask                   & $\mathcal{L}_{G}$                       & Generator loss                    \\
$\delta$          & Adversarial perturbation      & $\mathcal{L}_{tv}$                      & Total Variation loss              \\
$p$               & Adversarial patch or pattern  & $\mathcal{L}_{content}$                 & Content loss                      \\
$G$               & Generator                     & $\mathcal{L}_{blk}$                     & Black patch loss                  \\
$D$               & Discriminator                 & $\mathcal{L}_{style}$                   & Style loss                        \\
$z$               & Input noise                   & $\mathcal{L}_{nps}$                     & NPS loss                          \\
$\mathcal{T}^{a}$                       & Sticker transformation module
& $\mathcal{L}_{ssim}$                    & SSIM loss                         \\
$\mathcal{T}^{b}$ & Face transformation module & $\mathcal{L}_{lpips}$                   & LPIPS loss\\ \bottomrule
\end{tabular}}
\end{table}

\subsubsection{Technical terms}
\paragraph{Adversary's Knowledge}
\begin{itemize}
    \item \textit{White-box attack} means the attacker has full access to the model’s architecture, parameters, and gradients, enabling precise optimization of adversarial perturbations. However, this assumption is often impractical in real-world scenarios, where such access is typically restricted.
    \item \textit{Black-box attack} refers to a situation where the attacker can only query the target model using input samples and observe the outputs, without any internal access. This setup more closely reflects real-world conditions and presents greater challenges. Attackers often rely on transferring adversarial perturbations from a white-box surrogate model to execute attacks via transferability.
\end{itemize}

\paragraph{Adversarial Specificity}
\begin{itemize}
    \item \textit{Targeted attack} (\textit{impersonation attack}) generates perturbations by minimizing the distance between the features of the original and target image, causing misclassification as a specific individual (e.g., an authorized user), potentially leading to identity theft or unauthorized access.
    \item \textit{Non-targeted attack} (\textit{dodging attack}) evades recognition by maximizing the distance between the image’s features and the correct classification boundary, inducing misclassification to arbitrary identity except the correct one.
\end{itemize} 

\paragraph{Perturbation Universality}
\begin{itemize}
    \item \textit{Universal Attack} generates perturbations that can mislead diverse inputs, offering broad applicability. However, 
    these universal adversarial perturbations are hard to optimize, resulting in a lower attack success rate.
    \item \textit{Individual Attack} generates unique perturbations for each input, making them highly effective for the corresponding sample. The main drawback is that this approach lacks flexibility and must be repeated for each new input.
\end{itemize}

\subsection{Background}
\subsubsection{Face Recognition Systems}

\begin{figure}[ht]
\centering
\includegraphics[width=1\linewidth]{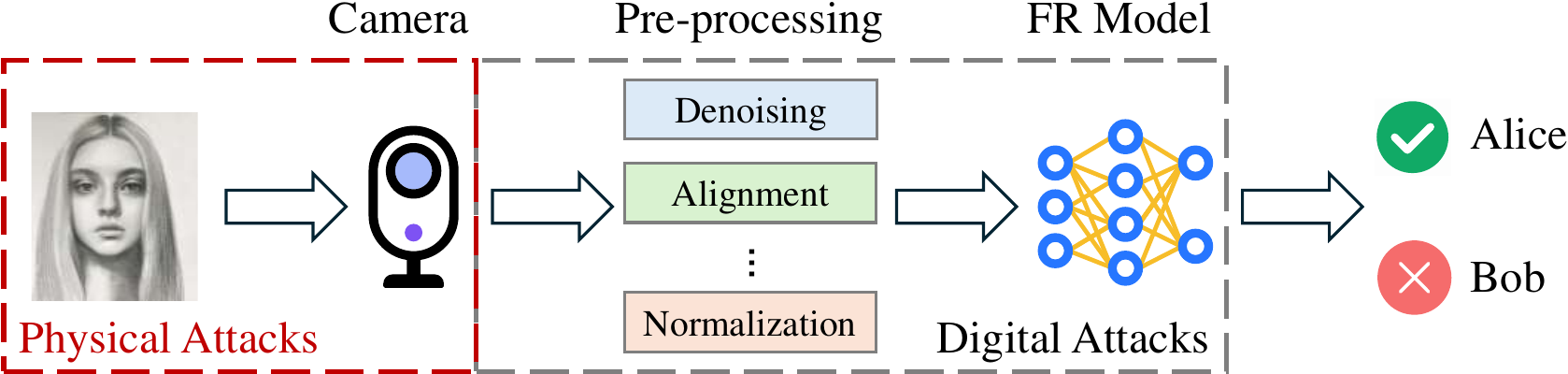}
\caption{Pipeline of the face recognition system. Physical attacks occur during the interaction between a person and the camera, while digital attacks directly operate on the digital pixels.}
\label{architecture}
\end{figure}

As illustrated in Fig. \ref{architecture}, a typical FR system consists of three key components: the camera, pre-processing module, and FR model. Specifically, a camera first captures the person's face in the physical world and converts it into a digital image, which is then processed by a pre-processing module that performs tasks such as denoising, alignment, and normalization to prepare the aligned face image for recognition. Finally, the processed image is input into an FR model, where feature vectors are extracted and used to identify the person based on their high-dimensional representations, completing the identification process. 

Adversarial attacks aim to introduce perturbations to face images, which cause the FR system to misidentify individuals, resulting in either a dodging attack (where no identity is recognized), or an impersonation attack (where the wrong identity is recognized). Digital adversarial attacks primarily target the pre-processing module and the FR model by injecting adversarial perturbations into the digital pixels of the image. In contrast, physical adversarial attacks occur during the interaction between a person and the camera in real-world environments, often carried out utilizing specialized physical mediums, such as adversarial glasses, hats, or stickers.

\subsubsection{Digital and Physical Adversarial Attacks}
Given a face image $x$ as input, the FR system recognizes its identity by matching features extracted by the feature extractor $f_\theta$. Digital adversarial attacks introduce perturbations to the pixel values of a digital image, resulting in an adversarial example $x^{adv}$ that appears visually similar to the original image $x$ but leads the FR model to make incorrect predictions, defined as:
\begin{equation}
    x^{adv} = x + \delta,
\end{equation}
where $\delta$ represents the global adversarial perturbations optimized by various digital attack techniques, such as PGD \cite{PGD}, L-BFGS \cite{szegedy2013intriguing}, MI-FGSM \cite{MI-FGSM}, C\&W \cite{CW}, and Deepfool \cite{Deepfool}.

Due to the difficulty of printing and capturing global perturbations of digital pixels in the physical world, physical attacks apply localized patches or patterns to the image. Such attacks are considered more practical and have attracted increasing attention from researchers. Physical adversarial attacks craft an adversarial sample $x^{adv}$ by adding adversarial perturbation $p$ to the benign image $x$, causing the FR system to fail in feature matching. $x^{adv}$ is defined as follows, with a binary mask $\mathcal{M}$ specifying the location and shape of $p$:
\begin{equation}
x^{adv}=(1-\mathcal{M})\cdot x+\mathcal{M}\cdot p.
\label{Eq:adversarial samples}
\end{equation}
where $p$ denotes an adversarial patch (such as a sticker, mask, hat, or glasses), or an adversarial pattern (like infrared, projection, and light applied to the face) in the physical world. 

\subsection{Benchmark}
We outline the benchmark for evaluating physical adversarial attacks against FR systems, including target models, benchmark datasets, evaluation settings in the physical world, and metrics to assess attack performance.

\subsubsection{Target Models}

\begin{figure*}[htbp] 
    \centering
    \scalebox{0.922}{ 
    \begin{tikzpicture}
        % Draw timeline
        \draw[->, very thick] (-0,0) -- (15.8,0);
        %画分界限
        % \draw[thick, dashed] (0,0) -- (0,4.2);
        % \draw[thick, dashed] (0,0) -- (0,-3.2);
        \draw[thick, dashed] (2,0) -- (2,4.2);
        \draw[thick, dashed] (2,0) -- (2,-3.2);
        \draw[thick, dashed] (4,0) -- (4,4.2);
        \draw[thick, dashed] (4,0) -- (4,-3.2);
        \draw[thick, dashed] (6,0) -- (6,4.2);
        \draw[thick, dashed] (6,0) -- (6,-3.2);
        \draw[thick, dashed] (8,0) -- (8,4.2);
        \draw[thick, dashed] (8,0) -- (8,-3.2);
        \draw[thick, dashed] (10,0) -- (10,4.2);
        \draw[thick, dashed] (10,0) -- (10,-3.2);
        \draw[thick, dashed] (12,0) -- (12,4.2);
        \draw[thick, dashed] (12,0) -- (12,-3.2);
        \draw[thick, dashed] (14,0) -- (14,4.2);
        \draw[thick, dashed] (14,0) -- (14,-3.2);
        % \draw[thick, dashed] (16,0) -- (16,4.2);
        % \draw[thick, dashed] (16,0) -- (16,-3.2);
        % Add years
        \node[below] at (1,0) {\small 2014};
        \node[below] at (3,0) {\small 2015};
        \node[below] at (5,0) {\small 2016};
        \node[below] at (7,0) {\small 2017};
        \node[below] at (9,0) {\small 2018};
        \node[below] at (11,0) {\small 2019};
        \node[below] at (13,0) {\small 2020};
        \node[below] at (15,0) {\small 2021};

        \node[above] at (1,3.5) {\large \textbf{2D}};
        \node[below] at (1,-2.5) {\large \textbf{3D}};

        % add 2D models
        \node[above] at (1,0.1) {\rotatebox{45}{\footnotesize DeepFace\cite{DeepFace}}};
        \node[above] at (1,0.7) {\rotatebox{45}{\footnotesize VGG16\cite{VGG16}}};
        \node[above] at (1,1.4) {\rotatebox{45}{\footnotesize Inception\cite{Inception}}};

        \draw[thick,->] (1.4,1.0) -- (2.4,1.0);
        \draw[thick,->] (1.2,2.2) -- (2.7,2.2);

        \node[above] at (3,0.4) {\rotatebox{45}{\footnotesize VGGFace\cite{VGGFace}}};
        \node[above] at (3,1.4) {\rotatebox{45}{\footnotesize FaceNet\cite{FaceNet}}};

        \draw[thick,->] (3.4,2.2) -- (4.4,1.5);
        \draw[thick,->] (1,2.5) -- (1,3) -- (1.8,3) -- (4.5,3);

        \node[above] at (5,0.1) {\rotatebox{45}{\footnotesize MTCNN\cite{MTCNN}}};
        \node[above] at (5,1.8) {\rotatebox{45}{\footnotesize \shortstack{Inception-\\ResNetV2\cite{Inception-ResNetV2}}}};
        \node[above] at (5,0.9) {\rotatebox{45}{\footnotesize OpenFace\cite{OpenFace}}};

        \node[above] at (7,0.1) {\rotatebox{45}{\footnotesize Faceboxes\cite{Faceboxes}}};
        \node[above] at (7,0.7) {\rotatebox{45}{\footnotesize SphereFace\cite{SphereFace}}};
        \node[above] at (7,1.4) {\rotatebox{45}{\footnotesize MobileNets\cite{MobileNets}}};

        \draw[thick,->] (7.6,2.5) -- (8.2,1.8);
        \draw[thick,->] (7.6,2.5) -- (8.2,2.5);

        \node[above] at (9,0) {\rotatebox{45}{\footnotesize PyramidBox\cite{Pyramidbox}}};
        \node[above] at (9,0.5) {\rotatebox{45}{\footnotesize LightCNN\cite{LightCNN29}}};
        \node[above] at (9,1.1) {\rotatebox{45}{\footnotesize ArcFace\cite{ArcFace}}};
        \node[above] at (9,1.3) {\rotatebox{45}{\footnotesize MobileFaceNets\cite{MobileFaceNets}}};
        \node[above] at (9,2.05) {\rotatebox{45}{\footnotesize MobileNetV2\cite{MobileNetV2}}};
        \node[above] at (9,2.8) {\rotatebox{45}{\footnotesize CosFace\cite{CosFace}}};
        
        \draw[thick,->] (9.6,2.2) -- (10.7,2.2);
        \draw[thick,->] (10.2,2.2) -- (10.2,2.8) -- (13,2.8) -- (13,1.5) ;
        \draw[thick,->] (13,2.8) -- (15,2.8) -- (15,1.7) ;

        \draw[thick] (9.7,4.0) -- (15,4.0) -- (15,2.8) ;

        \node[above] at (11,0.1) {\rotatebox{45}{\footnotesize DSFD\cite{DSFD}}};
        \node[above] at (11,0.6) {\rotatebox{45}{\footnotesize RetinaFace\cite{RetinaFace}}};
        \node[above] at (11,1.4) {\rotatebox{45}{\footnotesize AdaCos\cite{AdaCos}}};

        \node[above] at (13,0.1) {\rotatebox{45}{\footnotesize CurricularFace\cite{CurricularFace}}};

        \node[above] at (15,0.6) {\rotatebox{45}{\footnotesize MagFace\cite{MagFace}}};

        % add 3D models
        \node[below] at (7,-0.6) {\rotatebox{-45}{\footnotesize PointNet\cite{PointNet}}};
        \node[below] at (7,-1.5) {\rotatebox{-45}{\footnotesize PointNet++\cite{PointNet++}}};

        \draw[thick,->] (7,-1.6) -- (7,-2);
        \draw[thick,->] (7.2,-1.3) -- (8.7,-1.3);

        \node[below] at (9,-0.6) {\rotatebox{-45}{\footnotesize DGCNN\cite{DGCNN}}};
        \node[below] at (9,-1.4) {\rotatebox{-45}{\footnotesize 3DFaceNet\cite{3DFaceNet}}};

        \node[below] at (11,-0.6) {\rotatebox{-45}{\footnotesize FR3DNet\cite{FR3DNet}}};

        \node[below] at (15,-0.6) {\rotatebox{-45}{\footnotesize CurveNet\cite{CurveNet}}};
        
    \end{tikzpicture}
    }
    \caption{Timeline of advances and interconnections in 2D and 3D face recognition models.}
    \label{models} 
\end{figure*}
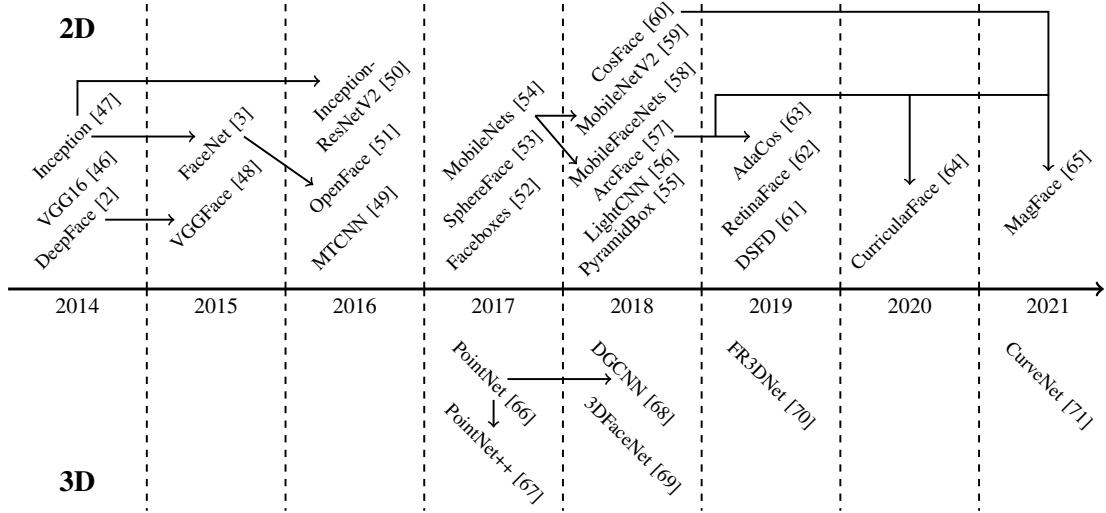
The study of security issues tends to lag, with existing physical adversarial attack methods focusing on classic models. 
Fig. \ref{models} illustrates the evolution and advancements of various architectures in FR models. These developments highlight the growing complexity and performance of 2D FR models, including diverse architectures such as Convolutional Neural Networks (CNNs) like ResNet\cite{ResNet} and Inception\cite{Inception}, lightweight models like MobileNet\cite{MobileNets} and MobileFaceNets\cite{MobileFaceNets}, and advanced frameworks incorporating angular-based losses like ArcFace\cite{ArcFace} and CosFace\cite{CosFace}. Moreover, 3D face recognition models such as PointNet\cite{PointNet} and DGCNN\cite{DGCNN} improve spatial feature handling.

\subsubsection{Datasets}

Benchmark datasets are crucial for evaluating the performance and generalization of physical adversarial attacks. As shown in Table \ref{tab:datasets}, these datasets are broadly categorized into two types: 2D and 3D face datasets. 2D Face Datasets provide large-scale collections of 2D face images that capture various variations in pose, lighting, expression, and demographics, essential for training and evaluating FR models against physical adversarial attacks. 3D Face Datasets introduce spatial information, enabling more accurate modeling of facial geometry and details like depth, pose, and expressions.

\subsubsection{Physical Evaluation Settings}
In addition to using the above models and datasets to generate adversarial samples and assess their performance in digital scenarios, it is essential to validate their effectiveness in real-world environments. Physical evaluations are typically conducted in two settings, determined by the level of experimental control: controlled laboratory settings or uncontrolled real-world settings.

In laboratory settings, researchers build experimental frameworks to simulate real FR systems, allowing for systematic control of environmental factors such as lighting, camera positions, and distances to evaluate the robustness of physical adversarial attacks\cite{patches, MTCNN-patches, noise, AdvHat, FaceAdv, GenAP, RHDE, Face3DAdv, PadvFace, wei2022simultaneously, DOPatch, SASMask, RSTAM, hwang2023adversarial, EAP, AdvSticker, AdvGlass, AGNs, glass-noise, cohen2023accessorize, singh2021brightness, CC2023, AdvMakeup, makeup2, AdvEye, ImU}. 
Specifically, attackers generate adversarial examples in the digital domain and use high-quality printers like the Canon SELPHY CP1300 or HP DeskJet 2677 to create physical adversarial artifacts, such as adversarial glasses, stickers, or masks. Subjects equipped with these physical artifacts are photographed by cameras such as the Canon EOS 650D, iPhone 11 Pro Max, Intel RealSense D415, or Logitech C270, and the captured images are then input into open-source models or commercial APIs like Face++\cite{Face++}, Aliyun\cite{Aliyun}, Microsoft Azure Face API\cite{Microsoft}, Clarifai\cite{Clarifai}, Baidu\cite{Baidu} and Tencent\cite{Tencent} for evaluation.

In contrast to controlled experimental frameworks in laboratory settings, real-world settings utilize end-to-end FR systems to evaluate the attack robustness in uncontrolled conditions, facing challenges from unpredictable environments such as varying illuminations, dynamic backgrounds, and diverse camera angles \cite{MAKEUP, mask, checkerboard, AT3D}. More precisely, subjects wear adversarial artifacts to test smartphone unlocking features on devices like the iPhone 11, Samsung Galaxy S10, and Xiaomi Redmi K20 Pro, commercial applications with built-in camera modules such as Alipay and B612.

\subsubsection{Metrics}
Existing studies have proposed various metrics to evaluate the performance of physical adversarial attacks against face recognition systems. Here, we provide detailed formulas and descriptions of commonly used metrics. The references accompanying each metric correspond to the papers that utilized them for evaluation.

\begin{itemize}
    \item \textit{Attack Success Rate (ASR)}\cite{MTCNN-patches, noise, checkerboard, FaceAdv, GenAP, RHDE, Face3DAdv, AT3D, wei2022simultaneously, DOPatch, SASMask, RSTAM, hwang2023adversarial, EAP, AdvSticker, AdvGlass, AGNs, glass-noise, cohen2023accessorize, singh2021brightness, CC2023, AdvMakeup, AdvEye, ImU, VLA, Projection, Optical, LIM} is the most commonly used metric to evaluate the effectiveness of physical adversarial attacks, which measures the proportion of successful misclassifications induced by adversarial samples on the target model:
    \begin{equation}
    \mathrm{ASR}=\frac{N_{\mathrm{success}}}{N},
    \end{equation}
    where $N_{\mathrm{success}}$ denotes the number of successful attacks, and $N$ represents the total number of adversarial examples. Similarly, other studies\cite{mask, MAKEUP, makeup2, IMA} also employ the Recognition Rate (1 - ASR) for evaluation.
        
    \item \textit{Cosine Similarity}\cite{patches, mask, AdvHat, Face3DAdv, PadvFace, EAP, cohen2023accessorize, AdvMakeup, LZP, Relighting} measures the similarity between the feature vectors of $\tilde{x}$ and the adversarial image $x^{adv}$. In impersonation attacks, $\tilde{x}$ represents the target image $x^t$, where a larger value indicates a more effective attack. While in dodging attacks, $\tilde{x}$ denotes the original image $x$, with a smaller value indicating a more successful attack:
    \begin{equation}    \mathrm{Cosine Similarity}=\frac{f_\theta(\tilde{x}) \cdot f_\theta(x^{adv})}{\|f_\theta(\tilde{x})\|\|f_\theta(x^{adv})\|}.
    \end{equation}
        
    \item \textit{Number of Queries}\cite{RHDE, wei2022simultaneously, DOPatch} counts the queries made to the target model to generate an adversarial example. This metric is particularly important in black-box attacks, where minimizing the number of queries improves the imperceptibility and efficiency of the attack.
    
    \item \textit{Mean Confidence Score (MCS)}\cite{EAP} quantifies the average confidence of the FR model when classifying adversarial examples, with higher values indicating that the model is overly confident in its incorrect predictions when misled by adversarial perturbations:
    \begin{equation}    \mathrm{MCS}=\frac1N\sum_{i=1}^N\text{confidence}(x^{adv}_i).
    \end{equation}

\end{itemize}

\section{Overview}
\label{section: overview}

\begin{table*}[htbp]
\centering
\caption{Summary of benchmark datasets used in face recognition.}
\label{tab:datasets}
\tabcolsep=0.08cm
\begin{tabular}{cccccl}
\toprule
\multicolumn{1}{c|}{\textbf{Task}}                                                & \textbf{Dataset}          &  \textbf{\# Image Size}   & \textbf{\# Sample}   & \multicolumn{1}{c|}{\textbf{\# Identity}} & \multicolumn{1}{c}{\textbf{Cited Literature}}              \\ \hline
\multicolumn{1}{c|}{\multirow{17}{*}{\textbf{2D}}}  & LFW\cite{LFW}               & 250$\times$250         & 13,233     & \multicolumn{1}{c|}{5,749}        & \cite{FaceAdv, GenAP, RHDE, Face3DAdv, AT3D, wei2022simultaneously, DOPatch, SASMask, RSTAM, hwang2023adversarial, EAP, AdvSticker, AdvGlass, MAKEUP, AdvMakeup, IMA, LZP, VLA, Projection}
 \\
\multicolumn{1}{c|}{}                                   & PubFig\cite{PubFig}            & -               & 58,797     & \multicolumn{1}{c|}{200}        & \cite{AdvGlass, AGNs, CC2023}
\\
\multicolumn{1}{c|}{}                                   & YMU\cite{YMU}               & -               & 604       & \multicolumn{1}{c|}{151}        & \cite{MAKEUP} \\
\multicolumn{1}{c|}{}                                   & CASIA NIR-VIS 2.0\cite{li2013casia}  & -               & 17,580    & \multicolumn{1}{c|}{725}        & \cite{cohen2023accessorize}\\
\multicolumn{1}{c|}{}                                   & CASIA-WebFace\cite{CASIA-WebFace}     & 96$\times$96           & 490,000    & \multicolumn{1}{c|}{15,000}      & \cite{patches, mask, AdvHat, RSTAM, cohen2023accessorize, ImU}\\
\multicolumn{1}{c|}{}                                   & CelebA\cite{CelebA}           & 218$\times$178         & 202,599    & \multicolumn{1}{c|}{10,177}      & 
\cite{MTCNN-patches, mask, GenAP, RHDE, Face3DAdv, AT3D, wei2022simultaneously, DOPatch, RSTAM, EAP, Relighting}
\\
\multicolumn{1}{c|}{}                                   & Wider Face\cite{WiderFace}      & -               & 32,203    & \multicolumn{1}{c|}{-}          & \cite{MTCNN-patches}\\
\multicolumn{1}{c|}{}                                   & CFP\cite{CFP}              & -               & 7,000      & \multicolumn{1}{c|}{500}        & \cite{SASMask} \\
\multicolumn{1}{c|}{}                                   & CASIA-FaceV5\cite{CASIA-FaceV5}   & 480$\times$640         & 2,500      & \multicolumn{1}{c|}{500}        & \cite{RSTAM}\\
\multicolumn{1}{c|}{}                                   & VGGFace2\cite{VGGFace2}    & -               & 3,310,000   & \multicolumn{1}{c|}{9,131}       & \cite{SASMask, glass-noise, singh2021brightness, ImU, Relighting}\\
\multicolumn{1}{c|}{}                                   & AgeDB\cite{AgeDB}      & -               & 16,488     & \multicolumn{1}{c|}{568}        & \cite{SASMask}\\
\multicolumn{1}{c|}{}                                   & MS-Celeb\cite{MS-Celeb}    & -               & 6,464,018 & \multicolumn{1}{c|}{94,682}      & \cite{mask}\\
\multicolumn{1}{c|}{}                                   & CelebA-HQ\cite{CelebA-HQ}    & 1024$\times$1024       & 30,000     & \multicolumn{1}{c|}{-}          & \cite{GenAP, Face3DAdv, AT3D, RSTAM, EAP}\\
\multicolumn{1}{c|}{}                                   & MS1M-V2\cite{ArcFace}      & -               & 5,800,000   & \multicolumn{1}{c|}{85,000}      &\cite{PadvFace, glass-noise} \\
\multicolumn{1}{c|}{}                                   & MT\cite{MT}            & 361$\times$361         & 3,834      & \multicolumn{1}{c|}{-}          & \cite{RSTAM, AdvEye}\\
\multicolumn{1}{c|}{}                                   & K-Face\cite{K-Face}      & -               & 32,400,000  & \multicolumn{1}{c|}{1,000}       & \cite{noise}\\
\multicolumn{1}{c|}{}                                   & LADN\cite{LADN}      & -               & 635       & \multicolumn{1}{c|}{-}          & \cite{AdvMakeup, AdvEye}\\ \hline

\multicolumn{1}{c|}{\multirow{3}{*}{\textbf{3D}}} %& BU-3DFE\cite{BU-3DFE}           & 2006 & 1040$\times$1329       & 2,500      & 100        & \\
& Bosphorus\cite{Bosphorus}    & 1600$\times$1200       & 4,666      & \multicolumn{1}{c|}{105}        & \cite{Optical} \\
\multicolumn{1}{c|}{}                                   & Eurecom\cite{Eurecom}      & 256$\times$256, 640$\times$480 & 936       & \multicolumn{1}{c|}{52}         & \cite{Optical} \\
\multicolumn{1}{c|}{}                                   & SIAT-3DFE\cite{SIAT-3DFE}     & -               & 32,000    & \multicolumn{1}{c|}{500}        & \cite{Optical}\\ 
\bottomrule
\end{tabular}
\end{table*}

To address \textbf{RQ1}, we first discuss the differences between physical and digital attacks, and then introduce the unique challenges of physical attacks as identified in the literature.

\subsection{Differences Between Physical and Digital Attacks}
% There is a significant distinction between physical and digital adversarial attacks, particularly in the context of FR systems. 
Adversarial attacks are generally categorized into digital and physical types, primarily depending on the environment they target\cite{wang2023adversarial, vakhshiteh2021adversarial, Surveillance}.
To fully understand the challenges posed by adversarial attacks against face recognition systems in the physical world, it is essential to understand the core differences between these two types of attacks.
Digital attacks, though excel in digital environments, often struggle to maintain their effectiveness in the physical world. Although physical adversarial attacks are more complex to implement, they can be effective in both physical and digital domains.
This disparity arises primarily from the differing workflows of physical and digital adversarial attacks. 
In particular, while digital adversarial attacks follow a ``virtual-to-virtual" process, physical adversarial attacks involve a ``virtual-to-real-to-virtual" process, introducing additional information loss. 
% \looseness=-1

Specifically, in the virtual-to-real phase, attackers transform digital adversarial perturbations into \textit{physical mediums} through various methods, such as printing adversarial patches or projecting adversarial patterns onto human faces. The technique in which the adversarial physical medium is employed can significantly affect the attack's effectiveness. For instance, a high-resolution color printer can better preserve adversarial information compared to a low-resolution printer, resulting in a more successful attack. The material on which adversarial samples are printed, such as paper or silk, can also impact the attack’s effectiveness, depending on the carrier's ability to retain ink. Additionally, the strobe characteristics of projectors can influence how well a camera captures adversarial information when a pattern is projected onto a face.

In the real-to-virtual phase, physical adversarial samples are captured by a real-world camera and then transformed into digital images through resampling.  
The quality of this transformation largely depends on the resolution and capabilities of the camera used; higher-resolution cameras can capture more detail, resulting in less information loss compared to lower-resolution cameras.
Moreover, external environmental factors, such as lighting, weather, viewing angles, and the distance between the camera and the target, can introduce noise and artifacts, which may degrade the quality of the captured image and reduce the effectiveness of the adversarial attack. 

To execute physical attacks on deployed FR systems in the real world, attackers design adversarial perturbations in the form of physical objects, such as masks, eyeglasses, or stickers. These physical adversarial samples are then captured by cameras and translated back into the digital domain for processing. In the following section, we discuss the challenges posed by such a ``virtual-to-real-to-virtual" process.

\subsection{Challenges in Physical Attacks}

\begin{table*}[htbp]
\caption{Overview of physical adversarial attacks against face recognition systems with different physical mediums.}
\centering
\label{tab:category} 
\begin{forest}
  % forked edges,
  for tree={
  grow=east,
  reversed=true,%increase counter-clockwise
  anchor=base west,
  parent anchor=east,
  child anchor=west,
  base=center,
  rectangle,
  draw=black,
  rounded corners,align=center,
  minimum width=3em,
  edge+={darkgray, line width=1pt},
  %  l sep+=2.5pt,
  %  s sep+=-5pt,
  inner xsep=4pt,
  inner ysep=1pt,
  text centered
  },
  where level=1{text width=5em,fill=yellow!10}{},
  where level=2{text width=5em,fill=pink!10}{},
  where level=3{yshift=0.26pt,fill=blue!10}{},
  where level=4{yshift=0.26pt,fill=orange!10}{},
  where level=5{yshift=0.26pt}{},
  [\textbf{Physical Adversarial}  \\ \textbf{Attacks}, text width=9em, fill=orange!20,
    [\textbf{Disguise-based}, text width=9em
        [Accessory, text width=5em,
            [AdvHat~\cite{AdvHat}/Kaziakhmedov \textit{et al.}~\cite{MTCNN-patches}/AdvMask~\cite{mask}/AT3D~\cite{AT3D} \\
            SASMask~\cite{SASMask}/AdvGlass\cite{AdvGlass}/AGNs~\cite{AGNs}/Singh \textit{et al.}~\cite{glass-noise}\\
            Cohen \textit{et al.}\cite{cohen2023accessorize}/Singh \textit{et al.}~\cite{singh2021brightness}/Cai \textit{et al.}~\cite{CC2023}/RSTAM~\cite{RSTAM}
            ]
        ]
        [Sticker, text width=5em,
            [Pautov \textit{et al.}\cite{patches}/Ryu \textit{et al.}~\cite{noise}/Wei \textit{et al.}~\cite{wei2022simultaneously}/EAP~\cite{EAP}/RHDE~\cite{RHDE}\\Face3DAdv\cite{Face3DAdv}/PadvFace~\cite{PadvFace}/DOPatch~\cite{DOPatch}/Hwang \textit{et al.}~\cite{hwang2023adversarial}\\
            Zhou \textit{et al.}~\cite{checkerboard}/FaceAdv~\cite{FaceAdv}/GenAP~\cite{GenAP}/AdvSticker~\cite{AdvSticker}\\  
            ]
        ]
        [Makeup, text width=5em
            [Guetta \textit{et al.}\cite{MAKEUP}/AdvMakeup~\cite{AdvMakeup}/Lin \textit{et al.}~\cite{makeup2}/AdvEye~\cite{AdvEye}/ImU~\cite{ImU}]
        ]
    ]
        [\textbf{Infrared-based}, text width=9em, 
            [Yamada \textit{et al.}\cite{invisible}/Yamada \textit{et al.}~\cite{yamada2013privacy}/FacePET~\cite{FacePET}/IMA~\cite{IMA}/LZP~\cite{LZP}, text width=32em,fill=blue!10]
        ]
          [\textbf{Illumination-based}, text width=9em
            [VLA\cite{VLA}/Nguyen \textit{et al.}~\cite{Projection}/Li \textit{et al.}~\cite{Optical}/Zhang \textit{et al.}~\cite{Relighting}/LIM~\cite{LIM}, text width=32em,fill=blue!10
            ]
          ]
]
\end{forest}
\end{table*}

Based on the aforementioned differences, physical adversarial attacks rely on physical mediums and exhibit unique characteristics. 
One basic attribute is imperceptibility. The imperceptibility of physical adversarial samples differs from that of digital samples. In digital attacks, adversarial noise is constrained by a small $\epsilon$-norm, making the adversarial samples nearly indistinguishable from benign examples. 
In contrast, physical adversarial examples face different constraints, allowing for slight but subtle visual discrepancies, often disguised as everyday items like hats or masks. Therefore, selecting an appropriate physical medium is an important yet challenging factor in enhancing the imperceptibility of such attacks.

Another crucial attribute is robustness. Robustness refers to how well physical adversarial samples maintain their attack effectiveness despite varying environmental noise. If physical samples require strict environmental conditions, their applicability and practical value are significantly reduced. Strong robustness ensures that these samples remain effective through virtual-to-real and real-to-virtual transformations, adapting to dynamic environmental changes. Thus, enhancing robustness is a key challenge in the development of physical attacks.

A third challenge is the complexity and cost associated with producing physical adversarial samples. Complexity refers to the difficulty in generating the optimal physical adversarial perturbations, while cost relates to the financial feasibility of their development. Physical-world attackers utilize various techniques to design adversarial samples, but those that are difficult and expensive to produce tend to be less practical and harder to implement in real-world attack scenarios.

The fourth attribute is universality, which refers to the attack's ability to remain effective across diverse input samples. Designing a universal attack that consistently performs well despite variations in input data is a significant challenge, requiring broad applicability and high adaptability.

The final attribute is transferability—the ability of physical adversarial samples to succeed in black-box environments. Physical attacks take place in complex, real-world settings, where variations in face recognition system architectures and parameters hinder attack success. Enhancing the black-box capability of these attacks is therefore a critical focus within the exploration of physical adversarial examples.

\subsection{Fundamental Approaches}

To address these challenges, researchers have developed various mediums, such as hats and masks, to execute the attacks. Accordingly, their efforts have focused on optimizing physical adversarial examples designed for these mediums, considering the aforementioned attributes.
Two primary strategies are employed to optimize $p$: the pixel space optimization strategy and the latent space optimization strategy. The pixel space optimization strategy perturbs pixel values to optimize the objective function during the generation of $p$. The latent space optimization strategy typically refers to the Generative Adversarial Network (GAN), which consists of a generator and a discriminator to learn target data distributions from latent space in a competitive way to generate high-quality $p$.

\noindent\textbf{Pixel Space Strategy.} The pixel space optimization strategy adjusts pixel values of $p$ using gradient descent to optimize the objective function $\mathcal{L}(x,p)$ \footnotemark. In dodging attacks, the goal is to maximize the difference between the features of $x^{adv}$ and $x$ extracted by $f$ making faces undetectable by face recognition systems. Its objective function is defined as:
\begin{equation}
\min_{p} 1- \mathcal{L}(x,p) = 1- \mathcal{L}_{adv}(f_\theta(x^{adv}),f_\theta(x)),
\label{Eq:opti-dodging}
\footnotetext{For brevity, in the following, we refer to $\mathcal{L}(x,p)$ as the Adversarial Total Loss and $\mathcal{L}_{adv}$ as the Adversarial Classifier  Loss.}
\end{equation}
where $\mathcal{L}_{adv}$ is the classification loss of the FR model. 

In contrast, the goal of impersonation attacks is to minimize the distance of features between $x^{adv}$ and the target image $x^{t}$, making the face recognition system identify $x^{adv}$ as the target identity. The objective function is defined as follows:
\begin{equation}
\min \mathcal{L}(x,p) = \mathcal{L}_{adv}(f_\theta(x^{adv}),f_\theta(x^{t})).
\label{Eq:opti-impersonation}
\end{equation}

\noindent\textbf{Latent Space Strategy.} Benefit from some advantages of GANs (such as their knowledge of facial structures or textures), some works use the generator of GAN to generate and achieve more realistic, coherent, or visually natural $p$. The GAN consists of two components: the generator ($G$), which maps input noise to data samples, and the discriminator ($D$), which differentiates between real and generated samples.

The latent space strategy utilizes the generator ($G$) to generate realistic $p$ from the input noise $z$, making them visually similar to real samples, using the following loss function:
\begin{equation}
\label{LG}
\begin{aligned}
    \mathcal{L}_{G}&=-\mathbb{E}_{z\sim\mathbb{Z}}[\log D(x^{adv})],\\
\mathrm{s.t.} \,\,\, x^{adv}&=(1-\mathcal{M})\cdot x+\mathcal{M}\cdot G(z).
\end{aligned}
\end{equation}

Accordingly, the objective function for dodging attack is:
\begin{equation}
\min \mathcal{L}(x,p) = \mathcal{L}_G - \lambda \mathcal{L}_{adv}(f_\theta(x^{adv}), f_\theta(x)),
\label{Eq:GANs-dodging}
\end{equation}
where $\lambda$ is a regularization parameter to balance the Generator Loss and the Adversarial Classifier Loss. The objective function for an impersonation attack is denoted as:
\begin{equation}
\min \mathcal{L}(x,p) = \mathcal{L}_G + \lambda \mathcal{L}_{adv}(f_\theta(x^{adv}), f_\theta(x^t)).
\label{Eq:GANs-impersonation}
\end{equation}

 The objectives of physical adversarial defense can be divided into two main aspects: enhancing the model's robustness and detecting adversarial attacks. Robustness ensures that the model maintains high classification or recognition accuracy even when subjected to physical attacks. The robustness objective aims to minimize the maximum loss under adversarial patches and can be mathematically expressed as follows:
\begin{equation}
    \min_\theta\max_p\mathcal{L}(f_\theta(x+p),f_\theta(x)).
\end{equation}

Detecting adversarial attacks, on the other hand, aims to identify whether an input is adversarial, differentiating between benign and adversarial samples. A formal detection approach models a function that classifies input $x$ as either adversarial (outputting 1) or non-adversarial (outputting 0). 
\begin{equation}
    \mathrm{Detect}(x)=\begin{cases}1,&\mathrm{if}\left\|\nabla_x\mathcal{L}_{adv}(f_\theta(x),y)\right\|>\tau\\0,&\mathrm{otherwise}\end{cases},
\end{equation}
where $\nabla_x\mathcal{L}_{adv}$ represents the gradient of the loss function, and $\tau$ is a predefined threshold. The input is flagged as potentially adversarial if the gradient value surpasses the threshold.

Recent research efforts have further addressed the aforementioned challenges by refining the two fundamental approaches.
In the following section, we present a taxonomy based on the physical mediums, highlighting how existing works within each category have evolved in overcoming these challenges.

\section{Physical Adversarial Attacks Against Face Recognition Systems} 
\label{section: attack}
To address \textbf{RQ2}, we first categorize physical adversarial attacks based on the medium and then analyze in detail how these methods overcome the aforementioned challenges. 

\subsection{Methodology}

\begin{figure*}[!t]
\begin{subfigure}{0.32\textwidth}
  \centering
  % include the first image
  \includegraphics[width=\linewidth]{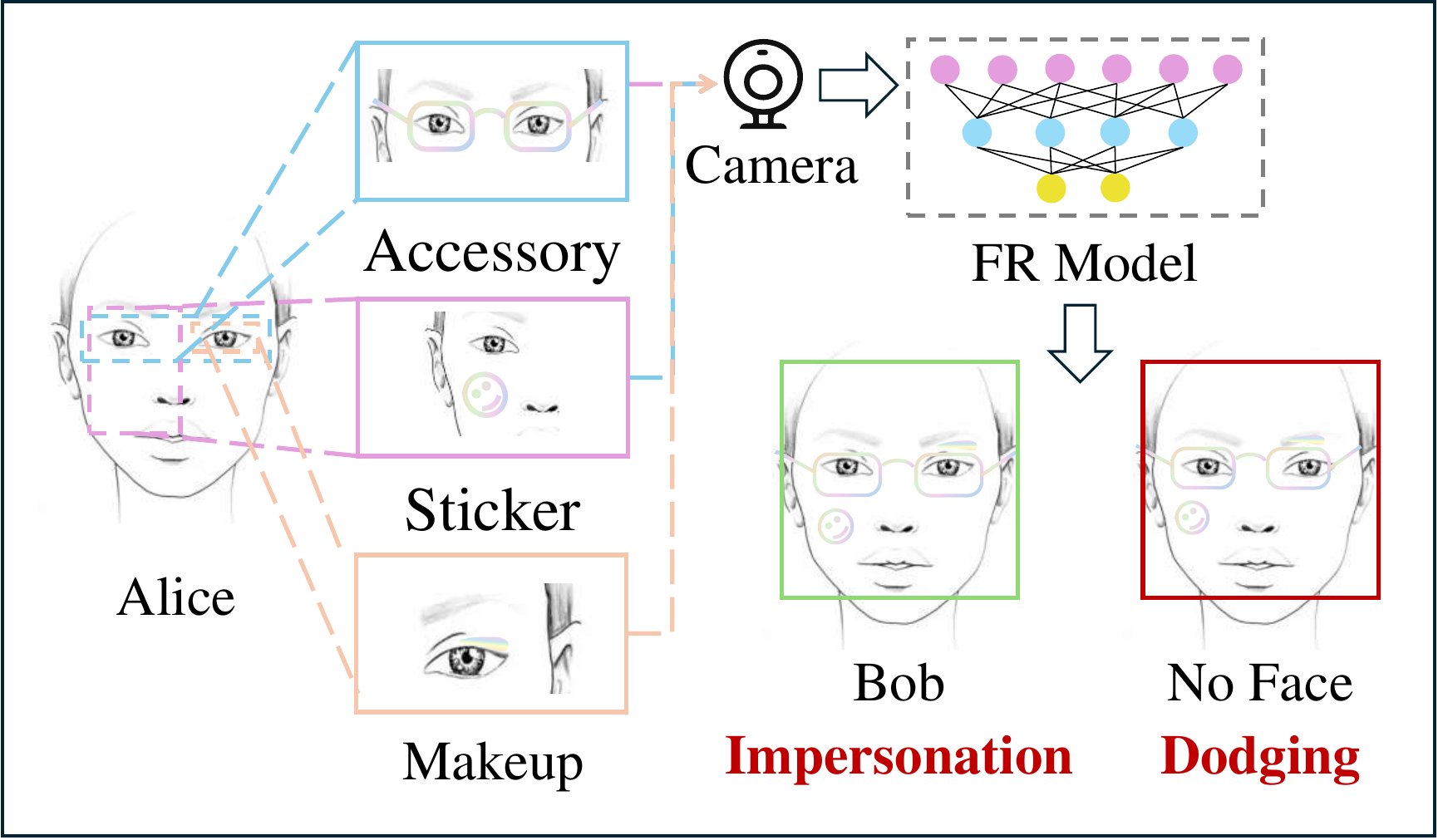}
  \caption{Adversarial attacks via disguises.}
  \label{face accessory}
\end{subfigure}
\hfil
\begin{subfigure}{0.32\textwidth}
  \centering
  \includegraphics[width=\linewidth]{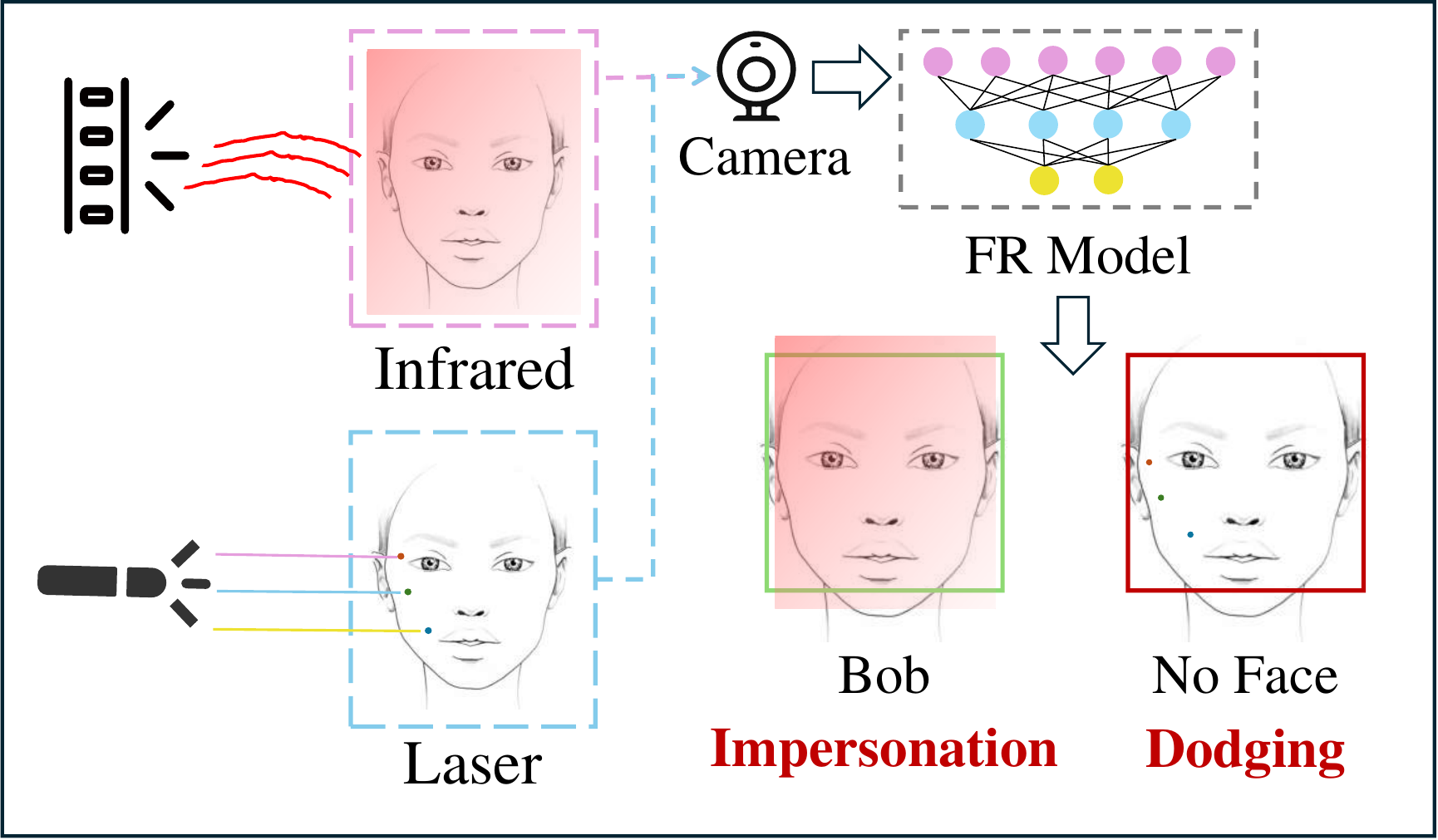}
  \caption{Adversarial attacks via infrared.}
  \label{infrared}
\end{subfigure}
\hfil
\begin{subfigure}{0.32\textwidth}
  \centering
  % include the second image
  \includegraphics[width=\linewidth]{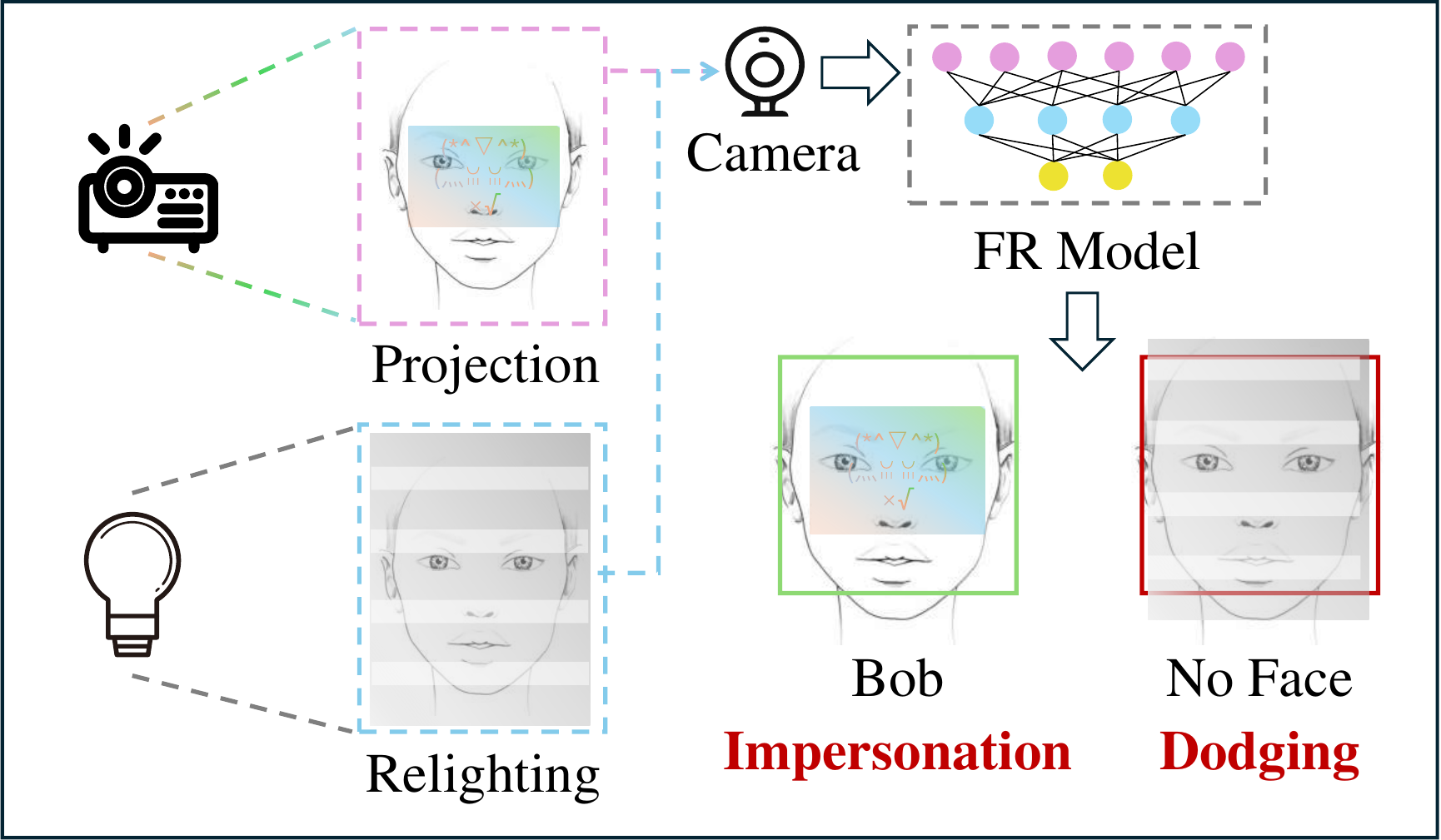}
  \caption{Adversarial attacks via illumination.}
  \label{projection}
\end{subfigure}
\caption{Illustration of three categories of physical adversarial attacks against face recognition systems, based on disguises, infrared, and illumination. The goal of these attacks is either to impersonate another individual or evade recognition.}
\label{Attack Schematic}
\end{figure*}
This section provides a detailed exploration of physical adversarial attacks on face recognition systems. We focus on the mediums used for physical adversarial attacks, attack attributes such as specificity and universality, and strategies including robustness, imperceptibility, and optimization. As shown in Table \ref{tab:category}, existing work can be classified into three categories based on the physical medium employed. The first category, disguise-based adversarial attacks, involves generating adversarial samples using items such as hats, masks, glasses, stickers, and makeup. These digitally-generated samples can be materialized through 3D printing for real-world application. The second category is infrared-based attacks, which utilize infrared or lasers. These attacks exploit the system's sensitivity to certain light wavelengths, employing infrared light or lasers to disrupt face recognition systems. Although imperceptible, their effectiveness is often limited by environmental conditions. The third category, illumination-based attacks, primarily involves projecting adversarial patterns onto the face or using point light sources to alter facial illumination. These methods allow for significant manipulation of lighting conditions and facial appearance. Fig. \ref{Attack Schematic} provides a schematic of three types of attacks, further detailed in Table \ref{tab:attacks}.

\begin{table*}[htbp]
\centering
\caption{Taxonomy of physical adversarial attacks against face recognition systems. ($\vartriangle$: dodging attacks, $\blacktriangle$: impersonation attacks; \Circle: white-box attacks, \CIRCLE: black-box attacks; $\square$: individual attacks, $\blacksquare$: universal attacks.)}
\label{tab:attacks} 
\tabcolsep=0.08cm
\begin{threeparttable}
{\begin{tabular}{cccccccc}
% {p{2cm}<{\centering}cp{1.1cm}<{\centering}p{1.1cm}<{\centering}p{2.6cm}<{\centering}p{1.1cm}<{\centering}p{2.8cm}<{\centering}p{2.5cm}<{\centering}p{2cm}<{\centering}}
\toprule
\textbf{Method} & \textbf{Medium} & \textbf{Specificity} & \textbf{Knowledge} & \textbf{Universality} & \textbf{Robustness} & \textbf{Imperceptibility} & \textbf{Optimization}\\
\midrule
   AdvHat\cite{AdvHat}& Hat & $\blacktriangle$ & \Circle & $\square$ & STL & TV & Pixel\\
   
   Kaziakhmedov \textit{et al.}\cite{MTCNN-patches}& Mask & $\vartriangle$ & \Circle & $\square$& EoT& TV & Pixel\\

   AdvMask\cite{mask}& Mask &  $\vartriangle$ & \CIRCLE & $\blacksquare$ & RLC-TA, UV & TV & Pixel\\
      % randomly apply location- and color-based transform augmentations
   RSTAM\cite{RSTAM}& Mask & $\blacktriangle$ & \CIRCLE & $\square$ & RST & - & Pixel \\

   AT3D\cite{AT3D}& Mask &$\blacktriangle$ & \CIRCLE & $\square$ & -& - & Pixel\\

   SASMask\cite{SASMask}& Mask & $\blacktriangle$ & \Circle & $\square$ & UV, EoT & SSIM, TV & Latent\\
   
   AdvGlass\cite{AdvGlass}& Glass & $\blacktriangle$ &\Circle & $\square$ & EoT, NPS &  TV & Pixel\\
   
   AGNs\cite{AGNs}& Glass &  $\blacktriangle$  & \Circle & $\blacksquare$ & EoT, NPS & - & Latent\\
   
   Singh \textit{et al.}\cite{glass-noise}& Glass & $\blacktriangle$ & \CIRCLE & $\square$  & - & TV & Pixel\\

   Cohen \textit{et al.}\cite{cohen2023accessorize}& Glass & $\blacktriangle$ & \Circle & $\square$  & EoT, NPS& TV & Pixel\\

   Singh \textit{et al.}\cite{singh2021brightness}& Glass &$\blacktriangle$ & \CIRCLE & $\square$  & - & - & Pixel\\

   Cai \textit{et al.}\cite{CC2023}& Glass &  $\blacktriangle$& \Circle & $\blacksquare$  & EoT, NPS & - & Latent\\ 

   Pautov \textit{et al.}\cite{patches} & Sticker & $\blacktriangle$ & \Circle & $\square$ & - & TV & Pixel\\

   Ryu \textit{et al.}\cite{noise}& Sticker & $\blacktriangle$ & \Circle & $\square$ & NPS & - & Pixel\\

   Zhou \textit{et al.}\cite{checkerboard}& Sticker & $\vartriangle$ & \CIRCLE & $\square$ & - & TV & Pixel\\

   FaceAdv\cite{FaceAdv}& Sticker & $\blacktriangle$ & \Circle & $\square$ & NPS& L2, LPIPS, SSIM, TV  & Latent\\

   GenAP\cite{GenAP}& Sticker &$\blacktriangle$ & \CIRCLE & $\square$ & - & - & Latent\\

   RHDE\cite{RHDE}& Sticker & $\blacktriangle$ & \CIRCLE & $\square$ & Real Sticker & Real Sticker & Pixel\\

   Face3DAdv\cite{Face3DAdv}& Sticker & $\blacktriangle$ & \Circle & $\square$ & - & - & Latent\\
   
   PadvFace\cite{PadvFace}& Sticker & $\blacktriangle$ & \Circle & $\square$ & EoT, D2P& - & Pixel\\

   Wei \textit{et al.}\cite{wei2022simultaneously}& Sticker & $\blacktriangle$ & \CIRCLE & $\square$ & - & TV & Pixel \\

   DOPatch\cite{DOPatch}& Sticker & $\blacktriangle$ & \CIRCLE & $\square$ & Real Sticker & Real Sticker & Pixel\\

   Hwang \textit{et al.}\cite{hwang2023adversarial}& Sticker & $\blacktriangle$ & \CIRCLE & $\square$ & - & - & Latent \\

   EAP\cite{EAP}& Sticker & $\blacktriangle$ & \CIRCLE & $\square$ & RST, IP& - & Pixel \\

   AdvSticker\cite{AdvSticker}& Sticker & $\blacktriangle$ & \CIRCLE & $\square$ & EoT & - & Pixel\\
   
   Guetta \textit{et al.}\cite{MAKEUP}& Makeup & $\vartriangle$ &\CIRCLE & $\square$ & - & - & Pixel \\

   AdvMakeup\cite{AdvMakeup}& Makeup & $\blacktriangle$ & \CIRCLE & $\square$ & - & MBS & Latent\\

   Lin \textit{et al.}\cite{makeup2}& Makeup &$\blacktriangle$ & \Circle & $\square$ & Gaussian blur& - & Latent\\
   
   AdvEye\cite{AdvEye}& Makeup & $\blacktriangle$ & \CIRCLE & $\square$ & Gaussian blur& LPIPS, SSIM & Latent\\

   ImU\cite{ImU}& Makeup & $\blacktriangle$ & \CIRCLE & $\square$ & StyleGANs & StyleGANs & Latent\\ \midrule

   Yamada \textit{et al.}\cite{invisible} & Infrared & $\vartriangle$ & \CIRCLE & $\blacksquare$ &  - & - &  Pixel \\
   
   IMA\cite{IMA} & Infrared  &$\blacktriangle$ & \Circle & $\square$ & -  & -  & Pixel \\
   
   Yamada \textit{et al.} \cite{yamada2013privacy}& Laser & $\vartriangle$ & \CIRCLE & $\square$ & - & -  & Pixel \\

   FacePET\cite{FacePET} & Laser & $\vartriangle$ & \CIRCLE & $\square$ & -  & -  & Pixel \\

   LZP\cite{LZP} & Laser & $\vartriangle$ & \CIRCLE & $\square$ & -  & -  & Pixel \\ \midrule

   VLA\cite{VLA} & Projection  &$\blacktriangle$ & \CIRCLE & $\square$ & Region\&Cluster  & POV   & Pixel  \\

   Nguyen \textit{et al.}\cite{Projection} & Projection & $\blacktriangle$ & \CIRCLE  & $\square$ & - & - & Pixel\\
   
   Li \textit{et al.}\cite{Optical} & Relighting & $\blacktriangle$ & \Circle  & $\square$  & 3D-TI&- & Pixel\\

   Zhang \textit{et al.}\cite{Relighting} & Relighting & $\vartriangle$ & \Circle  & $\square$ & - & - & Pixel\\

   LIM\cite{LIM} & Relighting  &$\vartriangle$ & \CIRCLE  & $\square$ & - & POV & Pixel\\
\bottomrule
\end{tabular}}
\begin{tablenotes}
\footnotesize
\item \textsuperscript{*} \textit{Notes: In general, methods to conduct impersonation attacks are also capable of performing dodging attacks.}
\end{tablenotes}
\end{threeparttable}
\end{table*}

Based on three types, we define seven evaluation criteria related to two categories: (1) effectiveness, which assesses the category's threat level, transferability, universality, robustness, and imperceptibility (criteria i-v); and (2) complexity, which evaluates the category's complexity of deployment and operation  (criteria vi-vii). To evaluate the performance of the three category attacks, the criteria are classified into three levels: low (\textit{L}), medium (\textit{M}), and high (\textit{H}). In some cases, a range of levels (e.g., low to medium (\textit{L-M}) or medium to high (\textit{M-H}) ) is used due to varying characteristics. 

(i) \textit{\textbf{Threat Level}} evaluates the potential harm attacks can cause in real-world scenarios. Impersonation attacks generally have a higher threat level than dodging attacks due to their ability to cause more severe consequences.

(ii) \textit{\textbf{Transferability}} assesses the attack effectiveness when the internal structure and parameters of the target system are unknown, crucial in real-world scenarios to assess the adaptability and practicality of the attack.

(iii) \textit{\textbf{Universality}} measures the attack effectiveness across different input samples, indicating universal applicability and high adaptability of the attack method.

(iv) \textit{\textbf{Robustness}} evaluates the consistency of the attack's performance under varying environmental conditions.    

(v) \textit{\textbf{Imperceptibility}} evaluates the imperceptibility of the attack, requiring a balance between it and effectiveness.

(vi) \textit{\textbf{Complexity}} assesses the difficulty of implementing and deploying adversarial attacks.
% considering the low level as Easy and the high level as Hard.

(vii) \textit{\textbf{Cost}} assesses the financial feasibility of developing an attack, as high costs can limit research accessibility.

\subsection{Disguise-based Adversarial Attacks}
As shown in Table \ref{tab:attacks}, we provide a detailed analysis of the performance of existing physical adversarial attack methods and how they address the aforementioned challenges, considering factors such as the physical medium, adversary's knowledge, attack specificity, robustness, imperceptibility strategies, and optimization techniques. Disguise-based attacks can be further classified into three subcategories depending on their specific implementation: adversarial accessories, adversarial stickers, and adversarial makeup. Adversarial masks, hats, and glasses fall under the accessory-based physical attacks category, as they print digital adversarial patches to be worn as face accessories. The other two commonly employed methods are adversarial stickers and makeup attacks.

Both the Pixel Space strategy and the Latent Space strategy can be employed to generate disguise-based adversarial attacks across different implementation subcategories. Thus, we organize the existing work into two main approaches—Pixel Space and Latent Space strategies—to provide a more structured introduction for each subcategory.

\begin{figure*}[t]
    \centering
    % 第一行的六个子图
    \begin{subfigure}[b]{0.15\textwidth}
        \centering
        \includegraphics[height=2.5cm]{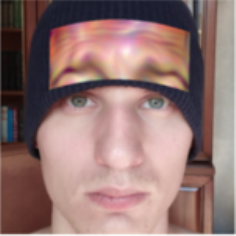} % 替换为你的图片路径
        \caption{AdvHat \cite{AdvHat}}
        \label{fig: AdvHat}
    \end{subfigure}
    \hspace{0.04cm}
    \begin{subfigure}[b]{0.15\textwidth}
        \centering
        \includegraphics[height=2.5cm]{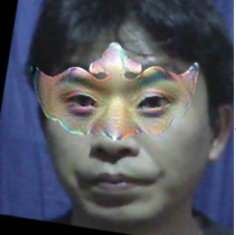}
        \caption{RSTAM\cite{RSTAM}}
        \label{fig: RSTAM}
    \end{subfigure}
    \hspace{0.06cm}
    \begin{subfigure}[b]{0.15\textwidth}
        \centering
        \includegraphics[height=2.5cm]{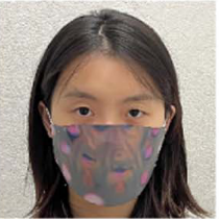}
        \caption{SASMask\cite{SASMask}}
        \label{fig: SASMask}
    \end{subfigure}
    \hspace{0.06cm}
    \begin{subfigure}[b]{0.15\textwidth}
        \centering
        \includegraphics[height=2.5cm]{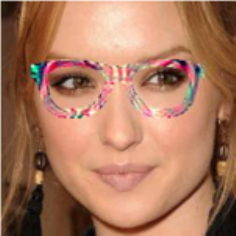}
        \caption{AdvGlass\cite{AdvGlass}}
        \label{fig: AdvGlass}
    \end{subfigure}
    \hspace{0.06cm}
    \begin{subfigure}[b]{0.15\textwidth}
        \centering
        \includegraphics[height=2.5cm]{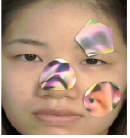}
        \caption{FaceAdv\cite{FaceAdv}}
        \label{fig: FaceAdv}
    \end{subfigure}
    \hspace{0.06cm}
    \begin{subfigure}[b]{0.15\textwidth}
        \centering
        \includegraphics[height=2.5cm]{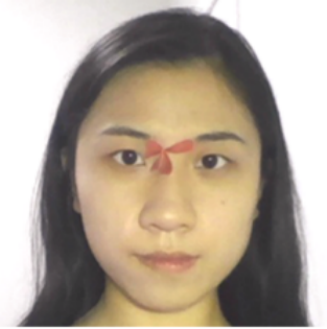}
        \caption{RHDE\cite{RHDE}}
        \label{fig: RHDE}
    \end{subfigure}

    \vspace{0.5cm} % 两行子图之间的垂直间距
    
    % 第二行的六个子图
    \begin{subfigure}[b]{0.15\textwidth}
        \centering
        \includegraphics[height=2.5cm]{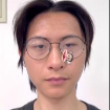}
        \caption{DOPatch\cite{DOPatch}}
        \label{fig: DOPatch}
    \end{subfigure}
    \hspace{0.06cm}
    \begin{subfigure}[b]{0.15\textwidth}
        \centering
        \includegraphics[height=2.5cm]{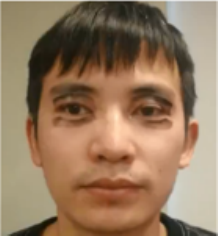}
        \caption{AdvMakeup\cite{AdvMakeup}}
        \label{fig: AdvMakeup}
    \end{subfigure}
    \hspace{0.06cm}
    \begin{subfigure}[b]{0.15\textwidth}
        \centering
        \includegraphics[height=2.5cm]{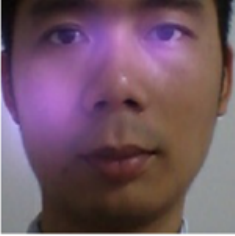}
        \caption{IMA\cite{IMA}}
        \label{fig: IMA}
    \end{subfigure}
    \hspace{0.06cm}
    \begin{subfigure}[b]{0.15\textwidth}
        \centering
        \includegraphics[height=2.5cm]{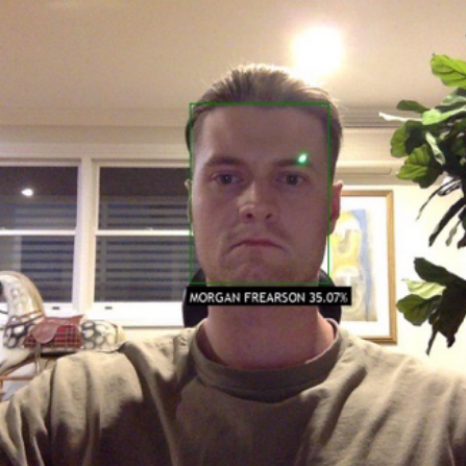}
        \caption{LZP\cite{LZP}}
        \label{fig: LZP}
    \end{subfigure}
    \hspace{0.06cm}
    \begin{subfigure}[b]{0.15\textwidth}
        \centering
        \includegraphics[height=2.5cm]{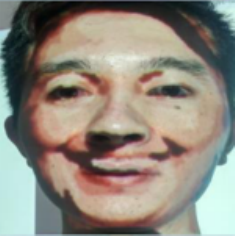}
        \caption{VLA\cite{VLA}}
        \label{fig: VLA}
    \end{subfigure}
    \hspace{0.06cm}
    \begin{subfigure}[b]{0.15\textwidth}
        \centering
        \includegraphics[height=2.5cm]{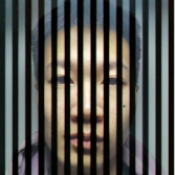}
        \caption{LIM\cite{LIM}}
        \label{fig: LIM}
    \end{subfigure}

    \caption{Examples of physical adversarial attacks against FR systems. \textbf{Disguise-based attacks:} (a) hat, (b-c) mask, (d) glass, (e-g) sticker, (h) makeup; \textbf{Infrared-based attacks:} (i) infrared, (j) laser; \textbf{Illumination-based attacks:} (k) projection, (l) relighting.}
\end{figure*}

\subsubsection{Accessory-based Adversarial Attacks}  \hspace{0.5em} \\  
% (\textbf{Threat Level}: \textit{M-H}, \textbf{Transferability}: \textit{L-M}, \textbf{Universality}: \textit{H}, \textbf{Robustness}: \textit{M-H}, \textbf{Imperceptibility}: \textit{M-H}, \textbf{Complexity}: \textit{L}, \textbf{Cost}: \textit{L})
\noindent\textbf{Pixel Space Strategy.} Eq. (\ref{Eq:opti-dodging}) and (\ref{Eq:opti-impersonation}) fulfill the fundamental objective of generating adversarial patches capable of deceiving FR models. However, they exhibit significant shortcomings in imperceptibility, robustness, and transferability.

To improve imperceptibility, AdvHat\cite{AdvHat} (see Fig. \ref{fig: AdvHat}) incorporated Total Variation (TV) loss \cite{tvloss} into the loss function to address the issue of over-perturbation and discontinuity between neighboring pixel values, and utilized Spatial Transformer Layer (STL) to map the patch on the face image. But they only carry out dodging attacks. As shown in Eq. (\ref{loss1}), $\lambda_{tv}$ serves as a regularization parameter that controls the influence of TV Loss. Adjusting $\lambda_{tv}$ can achieve a balance between adversarial effectiveness and image smoothness, allowing for deception without significantly altering visual perception.
\begin{equation}\label{loss1}\mathcal{L}(x,p)=\mathcal{L}_{adv}+\lambda_{tv} \mathcal{L}_{tv}.
\end{equation}

TV Loss measures pixel variation to ensure that the generated adversarial samples appear smoother, making them harder for the human eye to detect. It quantifies the total variation between neighboring pixels by calculating the square root of the sum of the squared differences between each pixel and its adjacent pixels to the right and below, as shown in Eq. \ref{TVloss}.
\begin{equation} 
\label{TVloss}\mathcal{L}_{tv}(p)=\sum_{i,j}\sqrt{(p_{i,j}-p_{i,j+1})^2+(p_{i,j}-p_{i+1,j})^2}.
\end{equation}

To further improve imperceptibility based on TV loss, Kaziakhmedov \textit{et al.}\cite{MTCNN-patches} integrated an optional black penalty loss $\mathcal{L}_{blk}$ to reduce an amount of black color on the patch enabling the mask to be less unusual (Eq. \ref{Eq: MTCNN-patches}).
\begin{equation}
\label{Eq: MTCNN-patches}
\mathcal{L}_{blk} = \sum_{i,j}1-p_{i,j}.
\end{equation}

In addition, Kaziakhmedov \textit{et al.} \cite{MTCNN-patches} employed the Expectation over Transformation (EoT) \cite{EOT} algorithm to enhance the robustness of physical adversarial samples, ensuring that adversarial patches remain effective in deceiving FR systems despite a variety of transformations. The EoT algorithm operates under the assumption that environmental factors such as camera angles, lighting conditions, and distance, affect the physical attack success. By simulating potential transformations $t$ from a set $T$, which includes rotation, scaling, cropping, lighting changes, and viewpoint variations, the EoT algorithm optimizes adversarial samples to maintain their efficacy across these transformations. Specifically, the EoT algorithm minimizes the expected loss across anticipated transformations, ensuring that the generated adversarial samples remain robust in both static and dynamic environments. The adversarial sample generated by the EoT algorithm is defined as follows:
\begin{equation}
x^{adv}=(1-\mathcal{M})\cdot x+\mathcal{M}\cdot \arg\min_p\mathbb{E}_{t\sim T}[\mathcal{L}_{adv}(f_\theta(x+p),f_\theta(x))].
\end{equation}

To improve robustness, AdvMask\cite{mask} built upon TV Loss by incorporating random location- and color-based transformations (RLC-TA) and digitally applying masks to face images using an end-to-end UV location map for universal dodging attacks. Additionally, an ensemble training approach using multiple face recognition models was adopted for black-box attacks. The UV position maps record the 3D coordinates of a complete facial point cloud from a 2D image, providing dense correspondence for each point in the UV space. This allows for a near-realistic approximation of the mask, essential for creating practical adversarial patches.

RSTAM \cite{RSTAM} (see Fig. \ref{fig: RSTAM}) enhanced the transferability of the adversarial masks through their proposed random similarity transformation (RST) strategy with four degrees of freedom (4DoF), which consists of translational, rotational, and scaling transformations. Furthermore, they proposed a random meta-optimization strategy for ensembling several pre-trained FR models to generate more universal adversarial masks that are highly effective in black-box impersonation attacks.

% Similar with RHDE\cite{RHDE}, 
To enhance the realism and adaptability of adversarial patches, particularly when viewed from different angles or under varying lighting conditions, researchers have turned to 3D models and rendering techniques. AT3D \cite{AT3D} utilized 3D Morphable Models (3DMM) \cite{3DMM} to manipulate the 3D geometry and texture of facial meshes. By incorporating neural rendering techniques \cite{pytorch3d}, AT3D generated highly natural 2D adversarial images from 3D models, which were more effective across different viewing angles and environmental conditions. Moreover, AT3D utilized surrogate models to improve the transferability of patches in black-box models.

AdvGlass \cite{AdvGlass}, a foundational study in adversarial glasses (see Fig. \ref{fig: AdvGlass}), introduced adversarial patches on eyeglass frames. It employed a loss function integrating TV loss for smooth color transitions and a non-printable score (NPS) loss to ensure printability. While it was not effective against state-of-the-art face recognition models, as demonstrated in AdvHat \cite{AdvHat}, its robustness strategy significantly influenced subsequent attack techniques. 
NPS, as shown in Eq. \ref{Eq: L_nps}, measures the quality degradation during the digital-to-physical transition and its impact on the sample’s overall effectiveness. Specifically, when an adversarial patch is printed on a physical medium (e.g., paper) and re-captured through a camera, the process introduces discrepancies due to factors like color deviation, brightness, and resolution limitations. These changes can affect the carefully controlled perturbations in the original digital patch, potentially weakening its adversarial effectiveness. Thus, NPS quantifies the degree of quality loss during the digital-to-physical transition and its impact on the overall effectiveness of the adversarial sample.
\begin{equation}
\label{Eq: L_nps}
\mathcal{L}_{nps}(p)=\sum_{\hat{p}\in G(z)}\left[\prod_{p\in\mathcal{P}}|\hat{p}-p|\right].
\end{equation}

Building on AdvGlass, Singh \textit{et al.}\cite{glass-noise} introduced the patch-noise combo attack, which combines the patch with imperceptibly small noises applied to other areas of the face image. They utilized regularization techniques to optimize TV loss and employed the Input Diversity Method\cite{xie2019improving}, Ensemble Diversity Method\cite{tramer2017ensemble}, or their combination to improve robustness and black-box transferability. Additionally, they incorporated the curriculum learning to enhance the resilience of adversarial samples to brightness variations, increasing their reliability in physical environments\cite{singh2021brightness}.
Further research explored adversarial glasses for specialized scenarios, including systems operating in near-infrared (NIR) wavelengths. Cohen \textit{et al.} \cite{cohen2023accessorize} adapted adversarial designs to this domain, incorporating strategies like TV loss, NPS, and EoT algorithm to maintain robustness, with optimization conducted through the PGD algorithm.
\begin{equation}
    \mathcal{L}(x,p)=\mathbb{E}_{t_1,t_2\thicksim T_1,T_2}\mathcal{L}_{adv}+\lambda_{tv}\mathcal{L}_{tv}+\lambda_{nps}\mathcal{L}_{nps},
\end{equation}
where $\mathbb{E}_{t_1,t_2\thicksim T_1,T_2}$ denotes the EoT transformation.

\noindent\textbf{Latent Space Strategy.} When dealing with FR systems, traditional adversarial attacks often focus on perturbing every pixel of generated patches to mislead the targeted system. Although somewhat effective, this strategy is limited by challenges in producing visually realistic adversarial samples and its potential ineffectiveness against complex or resistant systems. To address these challenges, more sophisticated strategies utilized GANs with distinctive generators and discriminators to learn target data distributions in latent space and generate highly realistic patches. 
The generator generated adversarial patches to deceive the FR system, while the discriminator continuously optimizes to distinguish between real and generated patches, thereby enhancing the generator’s ability to improve the patch quality. This recursive training process enables GANs to create adversarial patches that are both effective and imperceptible.

SASMask \cite{SASMask} (see Fig. \ref{fig: SASMask}) trained an adversarial style mask generator that hides adversarial perturbations inside style masks. Moreover, to ameliorate the phenomenon of sub-optimization with one fixed style, they proposed to discover the optimal style given a target through style optimization in a continuous relaxation manner. They simultaneously optimize the generator and the style selection for generating effective and stealthy adversarial style masks. Specifically, they introduced content loss $\mathcal{L}_{content}$ to remain the content of the source pattern and style loss $\mathcal{L}_{style}$ to generate the customized style. 
\begin{equation}
    \begin{aligned}&\mathcal{L}_{content}=\sum_{i\in l_C}\left\|\phi_i(M^{adv})-\phi_i(c)\right\|_2^2,\\&\mathcal{L}_{style}=\sum_{j\in l_{S}}\left\|\mathcal{G}(\phi_{j}(M^{adv}))-\mathcal{G}(\phi_{j}(s))\right\|_{2}^{2},\end{aligned}
\end{equation}
where $\phi_{l}(\cdot)$ is the output of the FR model at layer $l$, $l_C$ and $l_S$ are the sets of content and style layers; $\mathcal{G}(\phi_l(\cdot))$ is the Gram matrix; $c$ denotes the original mask pattern and $s$ represents the style of the predefined style image.  

AGNs \cite{AGNs} and Cai \textit{et al.} \cite{CC2023} utilized GANs to generate glasses capable of evading detection across various FR systems. These GANs-driven approaches demonstrated significant imperceptibility and effectiveness of adversarial glasses, laying the foundation for more advanced physical attacks.

\begin{remark}
Adversarial masks and glasses are suitable for different types of attacks. Masks are more effective for dodging attacks, as they obscure key facial features like the nose and mouth, making it difficult for FR systems to make accurate predictions but also limit the potential for modifying these regions' characteristics for impersonation attacks. In contrast, adversarial glasses are better suited for universal attacks because they specifically target the eye area, a consistent reference point in FR models. Their stable shape and adaptability make them effective across various faces. However, universal impersonation attacks, which require adversarial examples to mimic specific facial features across diverse individuals, are more complex and less explored in existing research.
\end{remark}

\subsubsection{Sticker-based Adversarial Attacks}

In contrast to accessories, which must be applied in fixed positions around faces, adversarial stickers offer greater flexibility. Their position and the perturbations of the patch can be dynamically optimized.

\noindent\textbf{Pixel Space Strategy.} Pautov \textit{et al.} \cite{patches} proposed a loss function similar to AdvHat's, utilizing grayscale patches to mitigate chromatic aberration from printers and cameras. This approach improves the robustness of adversarial stickers by minimizing color discrepancies during the real-to-virtual transformation process. AdvSticker\cite{AdvSticker} innovated by adding content loss $\mathcal{L}_{content}$ to minimize visual differences between the adversarial and benign image, facilitating black-box attacks through surrogate models with white-box access:
\begin{equation}
\label{AdvSticker}
\mathcal{L}_{content}=\sum_{i,j}\left(p_{i,j}-\frac1{N^2}\sum_{i=0}^N\sum_{j=0}^Nx_{i,j}\right)^2.
\end{equation}

Zhou \textit{et al.} \cite{checkerboard} demonstrated that the average classification loss across multiple FR models reveals common vulnerabilities in FR tasks. They utilized the total average classification loss function from multiple white-box models and TV loss as the optimization objective to enhance the transferability of the adversarial stickers in black-box attack scenarios. However, this method is limited to performing only dodging attacks. EAP \cite{EAP} further innovated by incorporating an image pyramid (IP) and meta-ensemble strategy into the RST framework. The strategies enabled the adaptation of adversarial patches to various scales and poses, as well as different camera angles and focal lengths, thus enhancing their effectiveness and transferability in the real-world environment.

Research\cite{noise, wei2022simultaneously, RHDE, DOPatch} have emphasized the importance of patch location in the effectiveness of adversarial attacks against FR systems. Researchers have explored the interplay between patch location and perturbation, demonstrating that carefully calibrated locations can significantly improve the attack success. Ryu \textit{et al.}\cite{noise} proposed three strategies for determining patch placement on the face: random selection, targeting locations that have the greatest impact on the FR system, and selecting areas with the highest noise concentration in adversarial examples, where only the facial region is modified. Similarly, Wei \textit{et al.}\cite{wei2022simultaneously} further refined the focus on patch location and perturbation as critical variables, introducing scale interpolation and TV Loss to enhance the smoothness of the patch's appearance. To extend these techniques to black-box settings, they leveraged ensemble attacks\cite{liu2016delving} using MI-FGSM\cite{MI-FGSM} along with reinforcement learning frameworks, effectively executing both dodging and impersonation attacks.

To improve imperceptibility and minimize information loss during the virtual-to-real transition, RHDE\cite{RHDE} (see Fig. \ref{fig: RHDE}) and DOPatch\cite{DOPatch} (see Fig. \ref{fig: DOPatch}) utilized real stickers to conduct attacks. 
RHDE found that the locations of the stickers on successful attacks show a regional aggregation. and employed a Heuristic Differential Evolutionary algorithm to reposition stickers based on output feedback from the model, without explicit gradients or model parameters. DOPatch\cite{DOPatch} introduced a distribution-optimization approach that efficiently searches for the worst-case regions for fixed-pattern patch placement, optimizing the distribution of adversarial locations across multiple points rather than focusing on individual locations. To address black-box constraints, DOPatch employed surrogate models to simulate the behavior of black-box systems.

PadvFace\cite{PadvFace} introduced the Digital to Physical (D2P) module $f_{d2p}$ to simulate virtual-to-real transformations, such as chromatic aberration. They also developed the Curriculum Adversarial Attack (CAA) algorithm to optimize adversarial patches across diverse physical conditions and improve robustness from virtual-to-real transformations. The adversarial sample is denoted as follows:
\begin{equation}
x^{adv}=\mathcal{T}^{b}((1-\mathcal{M})\odot x+\mathcal{M}\odot \mathcal{T}^{a}(f_{d2p}(p))),
\end{equation}
where $\mathcal{T}^{a}$ is a sticker transformation module, $\mathcal{T}^{b}$ is used to simulate environmental changes of faces such as different poses and lighting conditions.

\noindent\textbf{Latent Space Strategy.} Hwang \textit{et al.} \cite{hwang2023adversarial} and GenAP \cite{GenAP} utilized GANs to improve the transferability of adversarial patches of black-box models. GenAP constrained the patches to the low-dimensional manifold, enhancing their resemblance to natural facial features and facilitating the optimization process in black-box settings by surrogate model. Similarly, FaceAdv \cite{FaceAdv} (see Fig. \ref{fig: FaceAdv}) employed WGAN-GP \cite{WGAN-GP} to generate stickers to mimic real facial textures. To ensure the practicality of these stickers in the physical world, FaceAdv incorporated NPS loss to restrict colors within the printable RGB color space, maintaining visual fidelity after printing. 

Different from the conventional digital-to-physical transfer process, Face3DAdv \cite{Face3DAdv} utilized G3D's state-of-the-art pre-trained 3D generator \cite{G3D} to apply adversarial textures directly onto 3D face models and generated adversarial samples that account for various factors, enhancing the attack's realism and effectiveness in physical environments.

\begin{remark}
    Compared to methods that use accessories in fixed positions, sticker-based methods allow for the simultaneous optimization of both the patch location and the perturbation itself, which increases the attack's flexibility and expands the perturbation space. These attacks can deceive models with a single perturbation or combine multiple adversarial patches, further enhancing their effectiveness.
\end{remark}

\subsubsection{Makeup-based Adversarial Attacks}

In addition to strategies such as adversarial accessories and stickers, adversarial makeup has recently gained attention as a prominent area of research in physical attacks against face recognition systems. This approach manipulates facial features more subtly, blending adversarial modifications into natural cosmetic patterns to deceive FR systems while maintaining a realistic appearance.

\noindent\textbf{Pixel Space Strategy.} Guetta \textit{et al.}\cite{MAKEUP} optimized adversarial makeup on white-box surrogate models to deceive other black-box FR models. Specifically, during the generation of adversarial makeup in the digital domain, Guetta \textit{et al.} used a variant of the triplet loss function (\cite{FaceNet}) to compute a heatmap indicating the importance of facial features represented by the gradient strength, with higher gradients signifying greater importance. The adversarial makeup was then applied by a virtual makeup tool, You-Cam Makeup. They focused solely on dodging attacks and implemented the physical attack through manual makeup application by a makeup artist.

\noindent\textbf{Latent Space Strategy.} To improve the robustness and transferability of adversarial makeup, Lin \textit{et al.} \cite{makeup2} utilized Cycle-GAN \cite{Cycle-GAN} with Gaussian blur applied to the edges of the makeup. This technique helped soften the boundaries, allowing the adversarial makeup to blend more seamlessly with the natural facial features. The generated makeup was then manually applied by makeup artists. Cycle-GAN includes two GANs: one converts non-makeup images into makeup images, while the other reverses the process. Two discriminators ensure the realism of the generated makeup and the quality of the reconstructed images. ImU \cite{ImU} injected random noise into StyleGANs, which inherently lack any concept of facial identity, ensuring that the embedded latent vectors, despite representing the same physical person, do not have strong correlations. This approach helped achieve consistent adversarial makeup across varying poses and lighting conditions to improve robustness and imperceptibility. For optimization, ImU utilized gradient descent in white-box settings and genetic algorithms in black-box settings respectively.

To improve the imperceptibility, AdvMakeup \cite{AdvMakeup} (see Fig. \ref{fig: AdvMakeup}) applied makeup on eye regions, aiming to mislead FR models while remaining visually unnoticeable, appearing as natural makeup.
They employed the makeup blending strategy (MBS) to reduce style and content discrepancies between the source faces and the generated eye shadows, thereby enhancing the natural appearance of the adversarial faces. Furthermore, to improve attack transferability, the adversarial examples were designed to be model-agnostic. They introduced a fine-grained meta-learning attack strategy to generate better-generalized adversarial makeup, enhancing the transferability of the generated adversarial faces in black-box scenarios.

To improve the stability of manually applied adversarial makeup, AdvEye\cite{AdvEye} turned to simpler and more controllable methods, such as tattoo-like eyeshadow patches. They used GANs to generate natural eyeshadows with Gaussian blur to improve robustness and employed metrics like LPIPS (Learned Perceptual Image Patch Similarity) \cite{LPIPS} and SSIM (Structural Similarity Index Measure)\cite{SSIM} to enhance the imperceptibility of the attack. LPIPS is a deep-learning-based perceptual similarity metric that calculates the distance between images based on feature maps extracted from a pre-trained neural network (e.g., VGG \cite{VGG16} or AlexNet \cite{AlexNet}). Given two images $a$ and $b$, LPIPS extracts features $\phi$ and calculate distance:
\begin{equation}
\mathcal{L}_{lpips}(a,b)=\sum_l\frac1{H_lW_l}\sum_{h,w}||\phi_l(a)_{h,w}-\phi_l(b)_{h,w}||_2,
\end{equation}
where $\phi_l$ denotes the $l$-th layer feature map, and $H_l$ and $W_l$ are the height and width of the feature map.

SSIM is used to assess image quality, measuring the similarity of two images in brightness, contrast, and structure.
\begin{equation}
\mathcal{L}_{ssim}(a,b)=\frac{(2\mu_a\mu_b+C_1)(2\sigma_{ab}+C_2)}{(\mu_a^2+\mu_b^2+C_1)(\sigma_a^2+\sigma_b^2+C_2)},
\end{equation}
where $\mu_a$ and $\mu_b$ are the means of the images $a$ and $b$, $\sigma_a$ and $\sigma_b$ are the standard deviations, $\sigma_{ab}$ is the covariance, and $C_1$ and $C_2$ are small constants used for stabilization calculations.

To avoid overfitting to white-box models, AdvMakeup \cite{AdvMakeup} and AdvEye \cite{AdvEye} used the fine-grained meta-learning optimization strategy \cite{Meta-learning} to improve black-box transferability. 

\begin{remark}
    Latent space optimization strategies are particularly effective for makeup-based attacks. GANs exhibit significant strengths due to their strong feature learning capabilities in the latent space, supported by extensive makeup datasets. The interplay between the discriminator and generator enables GANs to produce highly realistic facial perturbations, such as subtle changes in skin tone, eyeliner, and lip color. 

\end{remark}

\subsection{Infrared-based Adversarial Attacks}

Another line of research claimed that disguise-based adversarial attacks might be relatively impractical. For example, while innovative, 3D-printed patches are highly conspicuous and likely to raise suspicion \cite{IMA}. To address this limitation,  recent studies utilized adversarial infrared light. Infrared-based methods aim to enhance the attack imperceptibility, as infrared light is invisible to the human eyes.

Early efforts by Yamada \textit{et al.} \cite{invisible} pioneered near-infrared LEDs in visors, which emitted invisible light detectable by sensors, distorting facial features used by FR models. Although this method demonstrated the potential for invisible attacks, it primarily targeted older face recognition techniques, limiting its effectiveness for more advanced FR systems. Subsequent research improved upon these early efforts. Moreover, Yamada \textit{et al.} \cite{yamada2013privacy} and FacePET \cite{FacePET} introduced lasers, where materials that absorbed and reflected light were applied to goggles, disrupting FR during preprocessing. These methods leveraged the reflective properties of light to undermine face detection, offering an alternative to infrared-based attacks. 

Further advancements, such as IMA \cite{IMA} (see Fig. \ref{fig: IMA}), utilized infrared LEDs mounted on wearable accessories like hats, optimizing their parameters to perform dodging and impersonation attacks. By fine-tuning the location, intensity, and size of infrared light points, these methods effectively disrupted the preprocessing stages of face recognition models, such as landmark extraction. LZP \cite{LZP} (see Fig. \ref{fig: LZP}) introduced a light pattern using a single laser point as an alternative to infrared-based attacks, which are often countered by solid-state devices like CMOS cameras with infrared cutoff filters. To enhance the effectiveness of this approach, LZP developed a light patch optimizer (LPO) to identify the optimal position for minimizing the confidence score of black-box FR systems, thereby improving the attack success rate.

\begin{remark}
    Most infrared-based attack methods are better suited for dodging attacks rather than impersonation attacks. 
    Impersonation attacks modify continuous facial features, such as the shape and contours of the face, to convincingly mimic another individual's identity. In contrast, dodging attacks require the alteration of discrete features to prevent the face recognition system from correctly identifying the individual.
    By targeting discrete points rather than continuous facial modifications, infrared-based attacks effectively hinder recognition without enabling the attacker to assume a different identity.
\end{remark}

\subsection{Illumination-based Adversarial Attacks} 
Adversarial patches are highly effective but lack stealth and are easily noticeable by humans. In contrast, infrared-based attack methods are more imperceptible but can be easily filtered by modern devices equipped with infrared filters. To balance effectiveness with imperceptibility, researchers have introduced a new strategy known as adversarial illumination.

Nguyen \textit{et al.}\cite{Projection} developed a transformation-invariant adversarial pattern generation method to generate a digital adversarial pattern projected onto the adversary’s face in the physical domain, demonstrating the vulnerability of face recognition systems to light projection attacks in both white-box and black-box attack settings.
\begin{equation}
x^{avg}=\mathcal{A}(x)=w_0x+\sum_{i=1}^{k-1}(w_i {T}_i(x)),s.t.\sum_{i=0}^{k-1}w_i=1,
\end{equation}
where ${T}_i$ corresponds to the $i$-th transformation, $k$ is the total number of transformations, and $w_i$ corresponds to the weight of ${T}_i$ such that $\sum_{i=0}^{k-1}w_{\boldsymbol{i}}=1$.

To improve imperceptibility, VLA \cite{VLA} (see Fig. \ref{fig: VLA}) used Persistence of Vision (POV) to create alternating frames that embedded adversarial patterns into face images, making them nearly invisible to humans while still disrupting FR systems. To further enhance robustness, VLA utilized a region-level perturbation strategy instead of pixel-level perturbation and applied clustering to group nearby similar colors into the same region. In contrast, LIM \cite{LIM} (see Fig. \ref{fig: LIM}) used high-speed on-off keying (OOK) to modulate light sources and generate rapid flickers, which were specifically designed for dodging attacks by exploiting the rolling shutter effect in CMOS sensors.

In contrast to the projection of adversarial patterns onto faces, Zhang \textit{et al.} \cite{Relighting} and Li \textit{et al.} \cite{Optical} focused on adversarial relighting, exploiting FR systems' sensitivity to specific lighting conditions. By fine-tuning the lighting direction and intensity, these methods made the attack process more natural and less detectable. Zhang \textit{et al.} developed a relighting strategy that used point light sources to simulate adversarial lighting conditions, which were physically replicated with a robotic arm for precise implementation. Li \textit{et al.} extended this concept to 3D face recognition systems, introducing structured light attacks that incorporated 3D reconstruction and skin reflectance models for effective adversarial relighting.

\begin{remark}
    Existing relighting-based and infrared-based attacks have not adopted latent space optimization strategies, as modifications in the latent space are difficult to map back to the light adjustment modification space. In contrast, projection-based attacks hold promise for generating adversarial patterns projected onto the face by leveraging latent space optimization strategies in future developments.
\end{remark}

\subsection{Summary}

To better understand the strengths, limitations, and practical implications of various attack mediums, we provide a summary of their evaluation here. Table \ref{tab:findings}  highlights that each category of adversarial attacks presents various trade-offs across factors such as threat level, transferability, universality, robustness, imperceptibility, complexity, and cost.

\begin{table}[htbp]
\centering
\caption{Comparison of different categories of physical adversarial attacks, with \textbf{Bold} indicating better performance.}
\label{tab:findings} 
\tabcolsep=0.05cm
\renewcommand{\arraystretch}{1.1}
\begin{threeparttable}
{ 
\begin{tabular}{cc|ccccccc}
\toprule

                          \multicolumn{2}{c|}{\textbf{Category}} & \textbf{\textit{Thr.}} & \textbf{\textit{Transf.}} & \textbf{\textit{Univ.}} & \textbf{\textit{Robust.}} & \textbf{\textit{Imper.}} & \textbf{\textit{Compl.}} & \textbf{\textit{Cost}} \\ \midrule
% \cline{1-9}
\multirow{3}{*}{{\textbf{Disguise}}}

& Accessory & M-H           & L-M                & \textbf{H}              & M-H           & M-H             & \textbf{L}              & \textbf{L}    \\
& Sticker & \textbf{H}          & M-H              & L              & \textbf{H}            & M            & \textbf{L}              & \textbf{L}    \\
& Makeup  & \textbf{H}             & \textbf{H}                & L  & L-M           & \textbf{H}               & M              & L-M  \\ \midrule
\textbf{Infrared}                  & -       & L-M           & \textbf{H}                & L                & L-M            & \textbf{H}               & H              & H    \\ \midrule
\textbf{Illumination}              & -       & M             & L                & L                & L-M            & M-H             & H              & H    \\ \bottomrule
\end{tabular}}
\begin{tablenotes}
\footnotesize
\item \textsuperscript{1} \textit{Thr. = Threat Level, Transf. = Transferability, Univ.  = Universality, Imper. = Imperceptibility, Compl. = Complexity.}
\item \textsuperscript{2} H = High, M = Medium, L = Low.
\end{tablenotes}
\end{threeparttable}
\end{table}

Several distinct strengths and limitations can be identified when focusing on disguise-based adversarial attacks, such as accessories, stickers, and makeup.
Sticker-based attacks offer a balanced threat level, making them a reliable choice for adversaries. Thanks to their strong universality and robustness, they are highly versatile and maintain their effectiveness across various conditions. Additionally, these attacks are cost-effective and straightforward to implement, which contributes to their widespread use. However, their noticeable nature can be a significant limitation, potentially reducing their efficacy in environments where imperceptibility is crucial. On the other hand, accessory-based attacks match sticker-based methods in terms of threat potential but have the advantage of being more imperceptible. They also demonstrate good adaptability across different samples due to their high universality. Despite these benefits. Makeup-based attacks stand out for their high adaptability and natural appearance, offering superior imperceptibility. They blend effectively with the user’s face, evading human detection more successfully than accessories or stickers. However, these attacks are more complex and resource-intensive to apply, making them challenging and costly to deploy. Additionally, their stability can be inconsistent, potentially impacting performance in varying conditions.

Infrared-based adversarial attacks demonstrate high black-box capability and effectively disrupt face recognition systems, as highlighted in recent research. By using infrared light, these attacks interfere with the camera's photoreceptors, impairing the system's ability to accurately detect facial features. This physical interference enables attackers to generate adversarial samples without the access to the system's internal parameters. However, despite their imperceptibility, infrared-based methods are often complex, costly, and limited by hardware requirements. Furthermore, as FR systems have become more advanced, particularly with the integration of infrared cutoff filters, the practicality of these methods has decreased.

Illumination-based attacks encounter even greater challenges. While projection-based methods can be useful in black-box settings, they are difficult to implement and achieve consistent results in physical environments. Although projections can mimic detailed facial features and spoof FR systems based on feedback, they require precise calibration of light sources and positions, leading to inconsistent results. In contrast, re-lighting attacks, typically employed in white-box settings, are even more challenging because they depend on the exact manipulation of light angles and intensities to alter facial features. The difficulty in controlling lighting accurately in real-world scenarios makes these attacks harder to execute.

Lastly, despite potential, infrared and illumination-based attacks remain under-researched, especially when compared to adversarial disguises. Several factors contribute to this gap. 
First, these attacks are highly sensitive to real-world variables, such as inconsistent lighting conditions, resulting in performance degradation when transitioning from digital simulations to physical implementations. Second, these attacks necessitate specialized equipment like projectors, infrared LEDs, or PWM circuits, which are difficult to manipulate and complex to set up. Moreover, manually calibrating light positions and precisely replicating experimental environments is challenging, resulting in high variability and making these methods difficult to reproduce. 
Lastly, as advanced FR systems have become more resistant to lighting changes, the effectiveness of these attacks has diminished. For instance, infrared-based attacks are often impractical since many cameras now include infrared cutoff filters, and ethical concerns such as potential harm to human eyes also limit their applicability.

In conclusion, while disguise-based methods including accessories, stickers, and makeup remain more popular and practical for adversarial attacks, infrared and illumination-based attacks, despite their strengths in imperceptibility, face significant implementation challenges and are less effective in real-world scenarios. Future research should focus on enhancing the stability and reducing the complexity of these methods to improve their viability for practical adversarial applications.

\subsection{Other Applications of Adversarial Samples in FR Systems}
Physical adversarial attacks act as a double-edged sword in FR systems. While they can deceive the system through minor patches, leading to incorrect identity verifications and enabling malicious users to bypass surveillance or impersonate others, they also offer benefits by protecting personal facial information from unauthorized surveillance using adversarial items like masks or glasses \cite{AdvGlass, invisible, yamada2013privacy, FacePET}.  Furthermore, Hasan \textit{et al.} \cite{hasan2023presentation}  demonstrated that these tactics not only disrupt FR models but also serve as defensive tools, allowing users to conceal their identities when needed. As FR technology advances, the role of adversarial attacks in privacy protection is expected to grow, driving research toward more effective, undetectable adversarial tools and stable privacy-preserving solutions across various environments.

\section{Adversarial Defense Against Physical Attacks}
\label{section: defense}

To address \textbf{RQ3}, we categorize and systematically analyze the existing physical adversarial defense methods.

Given the substantial threat posed by physical adversarial examples to real-world applications, researchers have focused on developing defenses to mitigate these risks. As shown in Table \ref{tab:defense}, this section reviews various defenses against physical adversarial attacks in FR systems. Unlike digital adversarial defenses, physical adversarial defenses are designed for real-world applications, targeting physically realizable attacks like local adversarial patch attacks. Specifically, the physical adversarial defense can be categorized into data-side and model-side defenses. Data-side defenses aim to eliminate adversarial influences or mitigate their effects, while model-side defenses work to improve the robustness of the model itself.

\begin{table}[htbp]
\centering
\caption{Summary of existing defense methods against physical adversarial attacks in face recognition systems.}
\label{tab:defense} 
\tabcolsep=0.08cm
\begin{tabular}{c|c|l}
\toprule
\textbf{Defense Paradigm}           & \textbf{Defense Sub-category} & \multicolumn{1}{c}{\textbf{Literature}}                                                                                                                                                                                                                                                                      \\ \midrule
\multirow{3}{*}{\makecell{Data-side \\ Defense}}  & Adversarial Detection         & \begin{tabular}[c]{@{}l@{}}
\cite{TaintRadar}, \cite{SentiNet}, \cite{xie2023random}, \cite{EAD}, \\ \cite{peng2024detection}, \cite{RADAP}, \cite{AmI}, \cite{UniFAD}
\end{tabular} \\ \cmidrule{2-3} 
                                    & Input Purification            & \cite{DIFFender}, \cite{PIN}, \cite{Jujutsu}                                                                                                                                                                                                                            \\ \midrule
\multirow{3}{*}{\makecell{Model-side \\ Defense}} & Adversarial Training          & \cite{DOA}, \cite{arvinte2020robust}, \cite{ren2024artificial}                                                                                                                                                                                                          \\ \cmidrule{2-3} 
                                    & Certified Defense             & \begin{tabular}[c]{@{}l@{}}\cite{chiang2020certified}, \cite{zhang2020clipped}, \cite{xiang2021detectorguard}, \cite{chen2022towards},\\ \cite{xiang2022patchcleanser}\end{tabular}                                                   \\ \bottomrule
\end{tabular}
\end{table}

\subsection{Data-side Defense}
The purpose of data-side defense is to mitigate the impact of adversarial patches, ensuring they do not compromise the face recognition system. In this paper, we categorize data-side adversarial defense into two classes: adversarial detection and input purification. Adversarial detection enables the FR system to identify and reject adversarial samples, preventing inaccurate or abnormal predictions. Input purification removes adversarial patches from incoming samples, restoring them to a benign state for accurate FR model prediction.

\subsubsection{Adversarial Detection}
Neural networks are highly sensitive to specific ``salient" regions in an image, meaning that any modification or removal of these areas can lead to significant classification errors. Adversarial examples exploit this sensitivity by targeting these key regions, causing drastic changes in the model's output, while benign examples remain relatively stable when these regions are altered. 

Neural networks are highly sensitive to specific ``salient" regions in an image, where even small modifications can lead to significant classification errors. Adversarial examples exploit this vulnerability by targeting these key regions, causing drastic shifts in the model's output. 
SentiNet\cite{SentiNet} leveraged this sensitivity as a detection mechanism, turning the vulnerability into an advantage. By utilizing model interpretability and object detection techniques, SentiNet effectively identified and detected localized universal adversarial patches. SentiNet pinpointed these salient regions, overlayed them on clean images, and tested for misclassifications. Regions likely to cause misclassifications were flagged as malicious, allowing for accurate detection of adversarial attacks.

Building upon SentiNet's approach, TaintRadar \cite{TaintRadar} further advanced the detection of localized adversarial examples. TaintRadar exploited the greater label variance in key regions of adversarial samples compared to benign ones, efficiently capturing complex localized attacks without requiring additional training or fine-tuning of the original model structure. This advancement addressed unique challenges that SentiNet did not fully resolve, enhancing the robustness of detection mechanisms against adversarial patches.
Xie \textit{et al.} \cite{xie2023random} proposed a random patch-based defense strategy that splitting an image into small patches can destroy the structure of an attack for the whole image. Once an attack was detected on an image patch, the whole image was considered to be attacked. Consequently, They segmented images into image blocks evenly or randomly. By training on these segmented image blocks, the model became more resilient to attacks affecting localized regions, effectively improving detection performance in both white-box and black-box scenarios.

Additionally, AmI \cite{AmI} proposed that adversarial samples exploit many neurons represented abstract features that are hard for humans to perceive, while benign inputs are classified based on human-perceptible features. AmI identified neurons critical to face attributes (e.g., eyes, nose) and used a novel bidirectional correspondence to establish strong correlations between human face attributes and internal neurons. By strengthening the values of critical neurons and weakening non-critical ones, AmI created a new model to enhance explainable reasoning and suppress unexplainable aspects. Inconsistent predictions between the original and transformed models indicated adversarial inputs. This technique did not require pre-knowledge of the attack or additional model training, making it robust against a wide range of attack vectors.

Inspired by active human perception and recurrent feedback mechanisms, EAD\cite{EAD} posited that environmental information can be actively contextualized to detect adversarial patches in the physical world.
Peng \textit{et al.} \cite{peng2024detection} extracted local face difference features from suspicious face images and reference face images to detect adversarial face accessories.
RADAP \cite{RADAP} employed FCutout and F-patch to improve the occlusion robustness of the face recognition model and the performance of the patch splitter using Fourier space sampling masks. In addition, an edge-aware binary cross entropy (EBCE) loss function is introduced to improve the accuracy of patch detection.
UniFAD\cite{UniFAD} employed a multi-task learning framework combined with k-means clustering to learn joint representations for coherent attacks, proposing a unified detection framework for three analogous threats: adversarial attacks, digital manipulation, and physical deception.

\subsubsection{Input Purification}
DIFFender\cite{DIFFender} was the first approach to utilize a text-guided diffusion model for defending against adversarial patches. By diffusing the adversarial examples with Gaussian noise and recovering the original inputs through the reverse denoising process, downstream classifiers could correctly recognize the denoised images with high robustness. 
However, it was observed that the patch region in the adversarial image could hardly be denoised towards the clean image. DIFFender identified adversarial patch regions by comparing differences across various denoised images and subsequently restored these regions while preserving the integrity of the underlying content. The method incorporated a text-guided diffusion model to localize and recover adversarial patches more accurately with textual prompts. 

There should exist a subspace called the immune space in which perturbations have fewer adverse effects on the recognition model than other subspaces. By analogy to the human immune system, a practical defense for face recognition is to inactivate adversarial perturbations, treating them as general noise, rather than improving model immunity. This strategy avoids overfitting by leveraging the model's inherent immunity without involving adversarial samples during training. Therefore, PIN\cite{PIN} was proposed to estimate the immune space and deactivate adversarial perturbations by restricting them to this subspace. A novel learnable PCA framework, inspired by PCA but adapted for nonlinearity, was used to train the neural network agent to select eigenvectors, effectively estimating the immune space and deactivating adversarial perturbations.
Patch attacks targeted only localized regions of the input, leaving most of the input intact. Consequently, Jujutsu\cite{Jujutsu} utilized GANs for local attack recovery by synthesizing the semantic content of corrupted inputs and reconstructing ``clean" inputs for accurate predictions. For detection, Jujutsu employed saliency maps to pinpoint adversarial patches, focusing on influential features. For purification, rather than masking the entire patch region, Jujutsu used uncorrupted pixels to reconstruct corrupted content with GANs, producing ``clean" images for accurate predictions. This method enhanced robust accuracy and reduced false positives on benign examples. 
\vspace{-8pt}

\subsection{Model-side Defense}
Model-side defenses aim to improve the model's inherent robustness, addressing its vulnerability to adversarial attacks. However, the ability of these defenses to generalize, especially against unknown attacks, remains uncertain. Model-side defense strategies can be categorized into adversarial training and certified defenses. Adversarial training mitigates adversarial vulnerability by incorporating adversarial examples into the model training process.  Certified defenses, on the other hand, provide theoretical robustness guarantees for deep models, offering users insight into the model's reliability.

\subsubsection{Adversarial Training}
DOA\cite{DOA} adopted adversarial training, similar to those in the digital domain, to defend against adversarial patches, leveraging the proposed Rectangular Occlusion Attack (ROA) for this purpose. This approach demonstrated significantly greater robustness against physical attacks: the eyeglass frame attack on FR systems and the sticker attack on stop signs.
Arvinte \textit{et al.} \cite{arvinte2020robust} employed contrast loss terms and untangled generative models to sample negative pairs during training, integrating them with adversarial training as an online enhancement method.
Renren \textit{et al.} \cite{ren2024artificial} drew inspiration from biological immune systems, which can recognize and respond to a wide range of threats, to introduce a self-supervised adversarial training mechanism that simulates immune system invasion.

\subsubsection{Certified Defense}
Certified defenses offer theoretical robustness guarantees for FR models. Given the guaranteed robustness bounds, model resistance to $L_p$-constrained attacks can be mathematically assessed. However, certified defenses have not been extensively applied in FR tasks, possibly due to the complexity of these models and the high dimensionality of facial data. We introduce them here to encourage future researchers to explore their potential in this context.

Chiang \textit{et al.} \cite{chiang2020certified} addressed the challenge of potentially stronger adversaries by training a robust model that provided a lower bound on adversarial accuracy and introduced the Interval Bound Propagation (IBP) defense. \cite{gowal2018effectiveness,mirman2018differentiable}, presenting the first certifiable defense against patch attacks and proposed modifications to IBP training to improve its efficiency in patch attacks. This certified patch defense was generalizable to patches of different shapes, and its robustness transferred well across various patches.
Zhang \textit{et al.} \cite{zhang2020clipped} introduced a scheme called clipped BagNet (CBN) to counteract adversarial stickers (patches), providing a certified security assurance for this method. 
Additionally, a strategy for provably securing object detectors was proposed by Xiang \textit{et al.} \cite{xiang2021detectorguard}, which involved an objectness explainer, relied on the mismatch judgment by an objectness explainer between the objectness predictor and the base detector. Addressing the same issue, Chen \textit{et al.} integrated the Vision Transformer architecture into the Derandomized Smoothing framework, resulting in superior certifiable patch defense with both high clean and certified accuracy \cite{chen2022towards}. Furthermore, Xiang \textit{et al.} proposed PatchCleanser, a double-masking defense framework, to achieve considerable certifiable robustness against adversarial patches in image classification\cite{xiang2022patchcleanser}. PatchCleanser effectively reduced dependency on specific model architectures, allowing for robust predictions even with adversarial inputs.

\section{Future Directions}
\label{section: future directions}
% \begin{tcolorbox}[size=title]
% \end{tcolorbox}

To address \textbf{RQ4}, we discuss potential future research directions, thereby inspiring relevant research in this area.

\vspace{-3pt}
\subsection{Benchmark Framework and Metrics}
Current research on physical adversarial attacks in FR systems faces significant challenges lacking a unified evaluation framework and standardized metrics. In particular, two major evaluation issues restrict accurate progress assessment.

\noindent\textbf{Differences in physical experimental settings:} Variations in physical conditions such as lighting, distance, angles, and environmental backgrounds make it challenging to compare methods fairly, even when using the same evaluation criteria. For instance, adversarial attacks effective in controlled laboratory settings with consistent lighting may perform poorly in real-world settings with variable lighting and backgrounds. These discrepancies hinder the accurate assessment and comparison of the effectiveness of different methods.

To address this, future efforts should focus on developing standardized physical test environments and protocols. Establishing a unified set of experimental setups spanning both controlled and real-world conditions would allow researchers to test their methods under comparable circumstances. Additionally, creating public datasets captured under these standardized conditions will facilitate fair comparisons across studies, allowing for more precise assessments of the strengths and weaknesses of various adversarial attack methods.

\noindent\textbf{Metric of imperceptibility} Assessing the imperceptibility of adversarial perturbations to the human eye remains challenging.  Current metrics such as L2 norm, LPIPS, SSIM, and PSNR primarily measure pixel-level differences or image quality, but they do not fully capture human perceptual factors in physical scenarios. Consequently, there is no widely accepted metric for physical imperceptibility, and most research omits quantitative assessment or depends on subjective visual evaluations, which lack standardization. 

Recent advancements in Visual Large Language models (VLLMs) offer promising avenues for automated perceptibility detection \cite{10445007}. By leveraging the capabilities of VLLMs, future research could develop automated systems that assess the perceptibility of adversarial perturbations based on human-like reasoning and contextual understanding. This approach would provide a standardized and scalable method for evaluating imperceptibility, reducing reliance on subjective assessments and enhancing the objectivity of evaluation metrics.

\vspace{-2pt}

\subsection{Advancing Attack Imperceptibility}
While several techniques have been developed to enhance attack imperceptibility, there remains room for further advancement in methods that effectively balance attack success with invisibility. 
Existing imperceptibility strategies in physical attacks often employ losses such as LPIPS, SSIM, PSNR, and TV loss, or utilize GANs to make perturbations less noticeable to the human. Building on these methods, future research could focus on designing perturbations that maintain high attack effectiveness and enhance imperceptibility.

\noindent\textbf{Integration of Multiple Mediums:} 
Combining various mediums, such as multiple types of jewelry or accessories, can achieve higher levels of imperceptibility. By distributing adversarial perturbations across different objects that naturally blend into everyday attire, attackers can create more subtle and less detectable adversarial examples. Future research could explore optimal combinations and placements of these mediums to maximize imperceptibility while ensuring attack efficacy.

\noindent\textbf{Leveraging Diffusion Models:} 
Integrating advanced generative models, particularly Diffusion Models like SDedit, with existing adversarial attack methods could significantly improve the imperceptibility of adversarial perturbations in the physical world. Recent advances in diffusion architectures have demonstrated superior performance in image generation and adversarial attack tasks within the digital domain of face recognition \cite{liu2024adv, hu2024towards}. Unlike GANs, diffusion models offer enhanced generation quality and model stability due to their gradual noise reduction process, which mitigates issues like mode collapse.
Building on these advancements, diffusion models hold significant potential for physical adversarial attacks, especially in the design of adversarial patches. Leveraging their strong generative capabilities, future research could focus on developing patches that are more invisible and harder to detect in physical environments.

\vspace{-3pt}

\subsection{Enhancing Attack Robustness}

Ensuring the robustness of physical adversarial attacks through virtual-to-real and real-to-virtual transformations is a critical aspect of their evaluation. Factors such as lighting conditions, angular deviations, and physical material properties significantly impact this transition. To address these challenges, future research should adopt a holistic approach by integrating existing techniques. For example, combining strategies like the EOT algorithm, which simulates diverse real-world scenarios, with advanced color gamut mapping and minimizing the NPS can enhance the effectiveness of perturbations when printed or displayed physically. Additionally, refining manual calibration alongside these methods can allow for precise adjustments necessary for practical deployment.

\vspace{-3pt}
\subsection{Advancing Black-Box capacity}
While existing black-box attack strategies have achieved notable success, they often rely on white-box surrogate models, which can lead to overfitting and reduced attack effectiveness when transferred to target models. Future research could focus on optimizing black-box attacks from two main perspectives: query-based and transferability-based strategies.

For query-based black-box attacks, developing more efficient methods for approximating gradient information and refining adversarial perturbations with fewer queries is crucial. This may utilize advanced optimization algorithms and reinforcement learning frameworks to enhance adversarial sample effectiveness while reducing interactions with the target model. For transferability-based black-box attacks, research could concentrate on improving the transferability of adversarial examples across different models. Ensemble learning and meta-learning have shown promise in generating robust adversarial samples with high attack success rates across various models. 
By advancing these areas, future research can enhance the effectiveness and efficiency of black-box attack strategies, enabling stronger attacks against diverse face recognition models.

\vspace{-3pt}
\subsection{Developing Universal Attacks}
Developing universal poses presents significant challenges, primarily because adversarial samples must consistently perform well across different faces and varying conditions. Although progress has been made, there is still considerable potential for improvement.
A key direction for future research is leveraging diffusion models for data augmentation to enhance the universality of adversarial attacks. Diffusion models can generate high-quality synthetic face data, significantly increasing the diversity and volume of training datasets. This enhanced diversity enables the adversarial patches to better adapt to various facial features and environmental conditions, improving their universality across different faces and scenarios. Furthermore, utilizing advanced network architectures and learning strategies can enhance the ability of adversarial attacks to generalize across unseen faces, increasing their effectiveness in real-world applications. By integrating diffusion models and simulating physical variations future research can develop more adaptable and universal adversarial attacks.

\vspace{-4pt}
\subsection{Optimizing Hyperparameters in Multi-Loss Functions}
Determining hyperparameters for multiple loss objective functions remains underexplored. Existing hyperparameter optimization approaches include several key methods. Grid Search evaluates model performance by exhaustively exploring all possible parameter combinations to identify the optimal set. While simple and intuitive, it is computationally expensive. Random Search, on the other hand, randomly samples parameter combinations within a predefined space and assesses their performance, offering greater efficiency than grid search, particularly in high-dimensional parameter spaces \cite{bergstra2012random}. Bayesian Optimization uses a prior probabilistic model, typically a Gaussian process, to predict the objective function and select new parameters that are likely to yield better results. By balancing exploration and exploitation, Bayesian optimization significantly reduces the number of parameter evaluations required \cite{snoek2012practical}. Additionally, Hyperparameter Optimization Libraries/Frameworks, such as Optuna and Hyperopt, provide flexible and efficient tools for automating hyperparameter searches. These libraries integrate multiple optimization algorithms, streamlining the search process \cite{akiba2019optuna}
% {, bergstra2013hyperopt}.

\vspace{-4pt}
\subsection{Strengthening Defense Mechanisms}
With the increasing complexity of physical adversarial attacks, relying solely on visual information may no longer provide sufficient defense. Future defense strategies could be strengthened by integrating multimodal data and combining inputs such as vision, depth information, infrared imaging, and sound. Additionally, multi-level biometric verification may enhance system security by incorporating other biometric methods, including fingerprint, iris, and voice recognition.

Incorporating simulated data of physical attacks into adversarial training of face recognition models could also improve the model's robustness to various environmental changes. Furthermore, employing generative models (e.g., GANs, Autoencoders, Diffusion Models) in detecting physical adversarial attacks shows promise. By using generative models, a distribution of ``normal" data can be established, allowing the system to detect inputs that significantly deviate from this distribution. If the deviation is substantial, the system could flag the input as a potential physical adversarial attack.

\vspace{-5pt}

\section{Conclusion}
\label{section: conclusion}

% This paper provides a systematic analysis of existing research literature on physical adversarial attacks targeting face recognition systems, aiming to stimulate further progress. 

In this paper, we provide a systematic analysis of existing research literature on physical adversarial attacks targeting face recognition systems. By identifying the unique challenges inherent in these attacks based on their specific workflows, we categorize them according to the physical medium used in real-world scenarios. We offer a detailed analysis of how existing attack methods address these challenges and summarize the strengths, limitations, and practical implications of each category. We examine current defenses against physical adversarial attacks. Moreover, we outline promising directions to facilitate future research. We hope that this paper deepens the understanding of key concepts and serves as a foundation for further advancing the security of face recognition systems.

\bibliographystyle{IEEEtran}
\bibliography{reference}

% Generated by IEEEtran.bst, version: 1.14 (2015/08/26)
\begin{thebibliography}{100}
\providecommand{\url}[1]{#1}
\csname url@samestyle\endcsname
\providecommand{\newblock}{\relax}
\providecommand{\bibinfo}[2]{#2}
\providecommand{\BIBentrySTDinterwordspacing}{\spaceskip=0pt\relax}
\providecommand{\BIBentryALTinterwordstretchfactor}{4}
\providecommand{\BIBentryALTinterwordspacing}{\spaceskip=\fontdimen2\font plus
\BIBentryALTinterwordstretchfactor\fontdimen3\font minus \fontdimen4\font\relax}
\providecommand{\BIBforeignlanguage}[2]{{%
\expandafter\ifx\csname l@#1\endcsname\relax
\typeout{** WARNING: IEEEtran.bst: No hyphenation pattern has been}%
\typeout{** loaded for the language `#1'. Using the pattern for}%
\typeout{** the default language instead.}%
\else
\language=\csname l@#1\endcsname
\fi
#2}}
\providecommand{\BIBdecl}{\relax}
\BIBdecl

\bibitem{lecun2015deep}
Y.~LeCun, Y.~Bengio, and G.~Hinton, ``Deep learning,'' \emph{nature}, vol. 521, no. 7553, pp. 436--444, 2015.

\bibitem{DeepFace}
Y.~Taigman, M.~Yang, M.~Ranzato, and L.~Wolf, ``Deepface: Closing the gap to human-level performance in face verification,'' in \emph{Proceedings of the IEEE conference on computer vision and pattern recognition}, 2014, pp. 1701--1708.

\bibitem{FaceNet}
F.~Schroff, D.~Kalenichenko, and J.~Philbin, ``Facenet: A unified embedding for face recognition and clustering,'' in \emph{Proceedings of the IEEE conference on computer vision and pattern recognition}, 2015, pp. 815--823.

\bibitem{patel2016secure}
K.~Patel, H.~Han, and A.~K. Jain, ``Secure face unlock: Spoof detection on smartphones,'' \emph{IEEE transactions on information forensics and security}, vol.~11, no.~10, pp. 2268--2283, 2016.

\bibitem{wang2021exploring}
J.~S. Wang, ``Exploring biometric identification in fintech applications based on the modified tam,'' \emph{Financial Innovation}, vol.~7, no.~1, p.~42, 2021.

\bibitem{goel2019development}
V.~Goel, H.~Raj, K.~Muthigi, S.~Sanjay~Kumar, D.~Prasad, and V.~Nath, ``Development of human detection system for security and military applications,'' in \emph{Proceedings of the Third International Conference on Microelectronics, Computing and Communication Systems: MCCS 2018}.\hskip 1em plus 0.5em minus 0.4em\relax Springer, 2019, pp. 195--200.

\bibitem{szegedy2013intriguing}
C.~Szegedy, W.~Zaremba, I.~Sutskever, J.~Bruna, D.~Erhan, I.~Goodfellow, and R.~Fergus, ``Intriguing properties of neural networks,'' in \emph{International Conference on Learning Representations}, 2014.

\bibitem{dong2019efficient}
Y.~Dong, H.~Su, B.~Wu, Z.~Li, W.~Liu, T.~Zhang, and J.~Zhu, ``Efficient decision-based black-box adversarial attacks on face recognition,'' in \emph{proceedings of the IEEE/CVF conference on computer vision and pattern recognition}, 2019, pp. 7714--7722.

\bibitem{zhong2020towards}
Y.~Zhong and W.~Deng, ``Towards transferable adversarial attack against deep face recognition,'' \emph{IEEE Transactions on Information Forensics and Security}, vol.~16, pp. 1452--1466, 2020.

\bibitem{yang2021attacks}
L.~Yang, Q.~Song, and Y.~Wu, ``Attacks on state-of-the-art face recognition using attentional adversarial attack generative network,'' \emph{Multimedia tools and applications}, vol.~80, pp. 855--875, 2021.

\bibitem{deb2020advfaces}
D.~Deb, J.~Zhang, and A.~K. Jain, ``Advfaces: Adversarial face synthesis,'' in \emph{2020 IEEE International Joint Conference on Biometrics (IJCB)}.\hskip 1em plus 0.5em minus 0.4em\relax IEEE, 2020, pp. 1--10.

\bibitem{jia2022adv}
S.~Jia, B.~Yin, T.~Yao, S.~Ding, C.~Shen, X.~Yang, and C.~Ma, ``Adv-attribute: Inconspicuous and transferable adversarial attack on face recognition,'' \emph{Advances in Neural Information Processing Systems}, vol.~35, pp. 34\,136--34\,147, 2022.

\bibitem{goswami2018unravelling}
G.~Goswami, N.~Ratha, A.~Agarwal, R.~Singh, and M.~Vatsa, ``Unravelling robustness of deep learning based face recognition against adversarial attacks,'' in \emph{Proceedings of the AAAI Conference on Artificial Intelligence}, vol.~32, no.~1, 2018.

\bibitem{zhong2019adversarial}
Y.~Zhong and W.~Deng, ``Adversarial learning with margin-based triplet embedding regularization,'' in \emph{Proceedings of the IEEE/CVF international conference on computer vision}, 2019, pp. 6549--6558.

\bibitem{chatzikyriakidis2019adversarial}
E.~Chatzikyriakidis, C.~Papaioannidis, and I.~Pitas, ``Adversarial face de-identification,'' in \emph{2019 IEEE International conference on image processing (ICIP)}.\hskip 1em plus 0.5em minus 0.4em\relax IEEE, 2019, pp. 684--688.

\bibitem{dabouei2019fast}
A.~Dabouei, S.~Soleymani, J.~Dawson, and N.~Nasrabadi, ``Fast geometrically-perturbed adversarial faces,'' in \emph{2019 IEEE Winter Conference on Applications of Computer Vision (WACV)}.\hskip 1em plus 0.5em minus 0.4em\relax IEEE, 2019, pp. 1979--1988.

\bibitem{FaceAdv}
M.~Shen, H.~Yu, L.~Zhu, K.~Xu, Q.~Li, and J.~Hu, ``Effective and robust physical-world attacks on deep learning face recognition systems,'' \emph{IEEE Transactions on Information Forensics and Security}, vol.~16, pp. 4063--4077, 2021.

\bibitem{GenAP}
Z.~Xiao, X.~Gao, C.~Fu, Y.~Dong, W.~Gao, X.~Zhang, J.~Zhou, and J.~Zhu, ``Improving transferability of adversarial patches on face recognition with generative models,'' in \emph{Proceedings of the IEEE/CVF conference on computer vision and pattern recognition}, 2021, pp. 11\,845--11\,854.

\bibitem{RHDE}
X.~Wei, Y.~Guo, and J.~Yu, ``Adversarial sticker: A stealthy attack method in the physical world,'' \emph{IEEE Transactions on Pattern Analysis and Machine Intelligence}, vol.~45, no.~3, pp. 2711--2725, 2022.

\bibitem{AdvSticker}
J.~Yang, Y.~Cheng, X.~Ji, and W.~Xu, ``A small sticker is enough: Spoofing face recognition systems via small stickers,'' in \emph{Proceedings of the 2022 6th International Conference on Electronic Information Technology and Computer Engineering}, 2022, pp. 1075--1082.

\bibitem{AdvGlass}
M.~Sharif, S.~Bhagavatula, L.~Bauer, and M.~K. Reiter, ``Accessorize to a crime: Real and stealthy attacks on state-of-the-art face recognition,'' in \emph{Proceedings of the 2016 acm sigsac conference on computer and communications security}, 2016, pp. 1528--1540.

\bibitem{AdvMakeup}
B.~Yin, W.~Wang, T.~Yao, J.~Guo, Z.~Kong, S.~Ding, J.~Li, and C.~Liu, ``Adv-makeup: A new imperceptible and transferable attack on face recognition,'' in \emph{Proceedings of the Thirtieth International Joint Conference on Artificial Intelligence, {IJCAI-21}}, Z.-H. Zhou, Ed.\hskip 1em plus 0.5em minus 0.4em\relax International Joint Conferences on Artificial Intelligence Organization, 8 2021, pp. 1252--1258.

\bibitem{MAKEUP}
N.~Guetta, A.~Shabtai, I.~Singh, S.~Momiyama, and Y.~Elovici, ``Dodging attack using carefully crafted natural makeup,'' \emph{arXiv preprint arXiv:2109.06467}, 2021.

\bibitem{AdvHat}
S.~Komkov and A.~Petiushko, ``Advhat: Real-world adversarial attack on arcface face id system,'' in \emph{2020 25th international conference on pattern recognition (ICPR)}.\hskip 1em plus 0.5em minus 0.4em\relax IEEE, 2021, pp. 819--826.

\bibitem{IMA}
Z.~Zhou, D.~Tang, X.~Wang, W.~Han, X.~Liu, and K.~Zhang, ``Invisible mask: Practical attacks on face recognition with infrared,'' \emph{arXiv preprint arXiv:1803.04683}, 2018.

\bibitem{invisible}
T.~Yamada, S.~Gohshi, and I.~Echizen, ``Use of invisible noise signals to prevent privacy invasion through face recognition from camera images,'' in \emph{Proceedings of the 20th ACM international conference on Multimedia}, 2012, pp. 1315--1316.

\bibitem{LZP}
M.~Frearson and K.~Nguyen, ``Adversarial attack on facial recognition using visible light,'' \emph{arXiv preprint arXiv:2011.12680}, 2020.

\bibitem{VLA}
M.~Shen, Z.~Liao, L.~Zhu, K.~Xu, and X.~Du, ``Vla: A practical visible light-based attack on face recognition systems in physical world,'' \emph{Proceedings of the ACM on Interactive, Mobile, Wearable and Ubiquitous Technologies}, vol.~3, no.~3, pp. 1--19, 2019.

\bibitem{Relighting}
Q.~Zhang, Q.~Guo, R.~Gao, F.~Juefei-Xu, H.~Yu, and W.~Feng, ``Adversarial relighting against face recognition,'' \emph{IEEE Transactions on Information Forensics and Security}, 2024.

\bibitem{LIM}
J.~Fang, C.~Jiang, Y.~Jiang, P.~Lin, Z.~Chen, Y.~Sun, S.-M. Yiu, and Z.~L. Jiang, ``Imperceptible physical attack against face recognition systems via led illumination modulation,'' \emph{IEEE Transactions on Big Data}, pp. 1--13, 2024.

\bibitem{FSG}
A.~Kurakin, I.~J. Goodfellow, and S.~Bengio, ``Adversarial examples in the physical world,'' in \emph{Artificial intelligence safety and security}.\hskip 1em plus 0.5em minus 0.4em\relax Chapman and Hall/CRC, 2018, pp. 99--112.

\bibitem{wang2022survey}
D.~Wang, W.~Yao, T.~Jiang, G.~Tang, and X.~Chen, ``A survey on physical adversarial attack in computer vision,'' \emph{arXiv preprint arXiv:2209.14262}, 2022.

\bibitem{wei2022visually}
X.~Wei, B.~Pu, J.~Lu, and B.~Wu, ``Visually adversarial attacks and defenses in the physical world: A survey,'' \emph{arXiv preprint arXiv:2211.01671}, 2022.

\bibitem{guesmi2023physical}
A.~Guesmi, M.~A. Hanif, B.~Ouni, and M.~Shafique, ``Physical adversarial attacks for camera-based smart systems: Current trends, categorization, applications, research challenges, and future outlook,'' \emph{IEEE Access}, 2023.

\bibitem{wang2023adversarial}
J.~Wang, D.~Wang, J.~Hu, S.~Wu, T.~Jiang, W.~Yao, A.~Liu, and X.~Liu, ``Adversarial examples in the physical world: A survey,'' \emph{arXiv preprint arXiv:2311.01473}, 2023.

\bibitem{wei2024physical}
H.~Wei, H.~Tang, X.~Jia, Z.~Wang, H.~Yu, Z.~Li, S.~Satoh, L.~Van~Gool, and Z.~Wang, ``Physical adversarial attack meets computer vision: A decade survey,'' \emph{IEEE Transactions on Pattern Analysis and Machine Intelligence}, 2024.

\bibitem{fang2024state}
J.~Fang, Y.~Jiang, C.~Jiang, Z.~L. Jiang, C.~Liu, and S.-M. Yiu, ``State-of-the-art optical-based physical adversarial attacks for deep learning computer vision systems,'' \emph{Expert Systems with Applications}, p. 123761, 2024.

\bibitem{kong2022digital}
C.~Kong, S.~Wang, H.~Li \emph{et~al.}, ``Digital and physical face attacks: Reviewing and one step further,'' \emph{APSIPA Transactions on Signal and Information Processing}, vol.~12, no.~1, 2022.

\bibitem{Surveillance}
K.~Nguyen, T.~Fernando, C.~Fookes, and S.~Sridharan, ``Physical adversarial attacks for surveillance: A survey,'' \emph{IEEE Transactions on Neural Networks and Learning Systems}, 2023.

\bibitem{hasan2023presentation}
M.~R. Hasan, R.~Guest, and F.~Deravi, ``Presentation-level privacy protection techniques for automated face recognition—a survey,'' \emph{ACM Computing Surveys}, vol.~55, no. 13s, pp. 1--27, 2023.

\bibitem{vakhshiteh2021adversarial}
F.~Vakhshiteh, A.~Nickabadi, and R.~Ramachandra, ``Adversarial attacks against face recognition: A comprehensive study,'' \emph{IEEE Access}, vol.~9, pp. 92\,735--92\,756, 2021.

\bibitem{PGD}
A.~Madry, A.~Makelov, L.~Schmidt, D.~Tsipras, and A.~Vladu, ``Towards deep learning models resistant to adversarial attacks,'' in \emph{International Conference on Learning Representations}, 2018.

\bibitem{MI-FGSM}
Y.~Dong, F.~Liao, T.~Pang, H.~Su, J.~Zhu, X.~Hu, and J.~Li, ``Boosting adversarial attacks with momentum,'' in \emph{Proceedings of the IEEE conference on computer vision and pattern recognition}, 2018, pp. 9185--9193.

\bibitem{CW}
N.~Carlini and D.~Wagner, ``Towards evaluating the robustness of neural networks,'' in \emph{2017 IEEE symposium on security and privacy (sp)}.\hskip 1em plus 0.5em minus 0.4em\relax Ieee, 2017, pp. 39--57.

\bibitem{Deepfool}
S.-M. Moosavi-Dezfooli, A.~Fawzi, and P.~Frossard, ``Deepfool: a simple and accurate method to fool deep neural networks,'' in \emph{Proceedings of the IEEE conference on computer vision and pattern recognition}, 2016, pp. 2574--2582.

\bibitem{VGG16}
K.~Simonyan and A.~Zisserman, ``Very deep convolutional networks for large-scale image recognition,'' in \emph{International Conference on Learning Representations}, 2015.

\bibitem{Inception}
C.~Szegedy, W.~Liu, Y.~Jia, P.~Sermanet, S.~Reed, D.~Anguelov, D.~Erhan, V.~Vanhoucke, and A.~Rabinovich, ``Going deeper with convolutions,'' in \emph{Proceedings of the IEEE conference on computer vision and pattern recognition}, 2015, pp. 1--9.

\bibitem{VGGFace}
O.~Parkhi, A.~Vedaldi, and A.~Zisserman, ``Deep face recognition,'' in \emph{BMVC 2015-Proceedings of the British Machine Vision Conference 2015}.\hskip 1em plus 0.5em minus 0.4em\relax British Machine Vision Association, 2015.

\bibitem{MTCNN}
K.~Zhang, Z.~Zhang, Z.~Li, and Y.~Qiao, ``Joint face detection and alignment using multitask cascaded convolutional networks,'' \emph{IEEE signal processing letters}, vol.~23, no.~10, pp. 1499--1503, 2016.

\bibitem{Inception-ResNetV2}
C.~Szegedy, S.~Ioffe, V.~Vanhoucke, and A.~Alemi, ``Inception-v4, inception-resnet and the impact of residual connections on learning,'' in \emph{Proceedings of the AAAI conference on artificial intelligence}, vol.~31, no.~1, 2017.

\bibitem{OpenFace}
B.~Amos, B.~Ludwiczuk, M.~Satyanarayanan \emph{et~al.}, ``Openface: A general-purpose face recognition library with mobile applications,'' \emph{CMU School of Computer Science}, vol.~6, no.~2, p.~20, 2016.

\bibitem{Faceboxes}
S.~Zhang, X.~Zhu, Z.~Lei, H.~Shi, X.~Wang, and S.~Z. Li, ``Faceboxes: A cpu real-time face detector with high accuracy,'' in \emph{2017 IEEE International Joint Conference on Biometrics (IJCB)}.\hskip 1em plus 0.5em minus 0.4em\relax IEEE, 2017, pp. 1--9.

\bibitem{SphereFace}
W.~Liu, Y.~Wen, Z.~Yu, M.~Li, B.~Raj, and L.~Song, ``Sphereface: Deep hypersphere embedding for face recognition,'' in \emph{Proceedings of the IEEE conference on computer vision and pattern recognition}, 2017, pp. 212--220.

\bibitem{MobileNets}
A.~Howard, ``Mobilenets: Efficient convolu-tional neural networks for mobile vision applications,'' \emph{arXiv preprint arXiv:1704.04861}, 2017.

\bibitem{Pyramidbox}
X.~Tang, D.~K. Du, Z.~He, and J.~Liu, ``Pyramidbox: A context-assisted single shot face detector,'' in \emph{Proceedings of the European conference on computer vision (ECCV)}, 2018, pp. 797--813.

\bibitem{LightCNN29}
X.~Wu, R.~He, Z.~Sun, and T.~Tan, ``A light cnn for deep face representation with noisy labels,'' \emph{IEEE transactions on information forensics and security}, vol.~13, no.~11, pp. 2884--2896, 2018.

\bibitem{ArcFace}
J.~Deng, J.~Guo, N.~Xue, and S.~Zafeiriou, ``Arcface: Additive angular margin loss for deep face recognition,'' in \emph{Proceedings of the IEEE/CVF conference on computer vision and pattern recognition}, 2019, pp. 4690--4699.

\bibitem{MobileFaceNets}
S.~Chen, Y.~Liu, X.~Gao, and Z.~Han, ``Mobilefacenets: Efficient cnns for accurate real-time face verification on mobile devices,'' in \emph{Biometric Recognition: 13th Chinese Conference, CCBR 2018, Urumqi, China, August 11-12, 2018, Proceedings 13}.\hskip 1em plus 0.5em minus 0.4em\relax Springer, 2018, pp. 428--438.

\bibitem{MobileNetV2}
M.~Sandler, A.~Howard, M.~Zhu, A.~Zhmoginov, and L.-C. Chen, ``Mobilenetv2: Inverted residuals and linear bottlenecks,'' in \emph{Proceedings of the IEEE conference on computer vision and pattern recognition}, 2018, pp. 4510--4520.

\bibitem{CosFace}
H.~Wang, Y.~Wang, Z.~Zhou, X.~Ji, D.~Gong, J.~Zhou, Z.~Li, and W.~Liu, ``Cosface: Large margin cosine loss for deep face recognition,'' in \emph{Proceedings of the IEEE conference on computer vision and pattern recognition}, 2018, pp. 5265--5274.

\bibitem{DSFD}
J.~Li, Y.~Wang, C.~Wang, Y.~Tai, J.~Qian, J.~Yang, C.~Wang, J.~Li, and F.~Huang, ``Dsfd: dual shot face detector,'' in \emph{Proceedings of the IEEE/CVF conference on computer vision and pattern recognition}, 2019, pp. 5060--5069.

\bibitem{RetinaFace}
J.~Deng, J.~Guo, E.~Ververas, I.~Kotsia, and S.~Zafeiriou, ``Retinaface: Single-shot multi-level face localisation in the wild,'' in \emph{2020 IEEE/CVF Conference on Computer Vision and Pattern Recognition (CVPR)}, 2020, pp. 5202--5211.

\bibitem{AdaCos}
X.~Zhang, R.~Zhao, Y.~Qiao, X.~Wang, and H.~Li, ``Adacos: Adaptively scaling cosine logits for effectively learning deep face representations,'' in \emph{Proceedings of the IEEE/CVF Conference on Computer Vision and Pattern Recognition}, 2019, pp. 10\,823--10\,832.

\bibitem{CurricularFace}
Y.~Huang, Y.~Wang, Y.~Tai, X.~Liu, P.~Shen, S.~Li, J.~Li, and F.~Huang, ``Curricularface: adaptive curriculum learning loss for deep face recognition,'' in \emph{proceedings of the IEEE/CVF conference on computer vision and pattern recognition}, 2020, pp. 5901--5910.

\bibitem{MagFace}
Q.~Meng, S.~Zhao, Z.~Huang, and F.~Zhou, ``Magface: A universal representation for face recognition and quality assessment,'' in \emph{Proceedings of the IEEE/CVF conference on computer vision and pattern recognition}, 2021, pp. 14\,225--14\,234.

\bibitem{PointNet}
C.~R. Qi, H.~Su, K.~Mo, and L.~J. Guibas, ``Pointnet: Deep learning on point sets for 3d classification and segmentation,'' in \emph{Proceedings of the IEEE conference on computer vision and pattern recognition}, 2017, pp. 652--660.

\bibitem{PointNet++}
C.~R. Qi, L.~Yi, H.~Su, and L.~J. Guibas, ``Pointnet++: Deep hierarchical feature learning on point sets in a metric space,'' \emph{Advances in neural information processing systems}, vol.~30, 2017.

\bibitem{DGCNN}
Y.~Wang, Y.~Sun, Z.~Liu, S.~E. Sarma, M.~M. Bronstein, and J.~M. Solomon, ``Dynamic graph cnn for learning on point clouds,'' \emph{ACM Transactions on Graphics (tog)}, vol.~38, no.~5, pp. 1--12, 2019.

\bibitem{3DFaceNet}
Y.~Guo, J.~Cai, B.~Jiang, J.~Zheng \emph{et~al.}, ``Cnn-based real-time dense face reconstruction with inverse-rendered photo-realistic face images,'' \emph{IEEE transactions on pattern analysis and machine intelligence}, vol.~41, no.~6, pp. 1294--1307, 2018.

\bibitem{FR3DNet}
S.~Z. Gilani and A.~Mian, ``Learning from millions of 3d scans for large-scale 3d face recognition,'' in \emph{Proceedings of the IEEE conference on computer vision and pattern recognition}, 2018, pp. 1896--1905.

\bibitem{CurveNet}
T.~Xiang, C.~Zhang, Y.~Song, J.~Yu, and W.~Cai, ``Walk in the cloud: Learning curves for point clouds shape analysis,'' in \emph{Proceedings of the IEEE/CVF International Conference on Computer Vision}, 2021, pp. 915--924.

\bibitem{ResNet}
K.~He, X.~Zhang, S.~Ren, and J.~Sun, ``Deep residual learning for image recognition,'' in \emph{Proceedings of the IEEE conference on computer vision and pattern recognition}, 2016, pp. 770--778.

\bibitem{patches}
M.~Pautov, G.~Melnikov, E.~Kaziakhmedov, K.~Kireev, and A.~Petiushko, ``On adversarial patches: real-world attack on arcface-100 face recognition system,'' in \emph{2019 International Multi-Conference on Engineering, Computer and Information Sciences (SIBIRCON)}.\hskip 1em plus 0.5em minus 0.4em\relax IEEE, 2019, pp. 0391--0396.

\bibitem{MTCNN-patches}
E.~Kaziakhmedov, K.~Kireev, G.~Melnikov, M.~Pautov, and A.~Petiushko, ``Real-world attack on mtcnn face detection system,'' in \emph{2019 International Multi-Conference on Engineering, Computer and Information Sciences (SIBIRCON)}.\hskip 1em plus 0.5em minus 0.4em\relax IEEE, 2019, pp. 0422--0427.

\bibitem{noise}
G.~Ryu, H.~Park, and D.~Choi, ``Adversarial attacks by attaching noise markers on the face against deep face recognition,'' \emph{Journal of information security and applications}, vol.~60, p. 102874, 2021.

\bibitem{Face3DAdv}
X.~Yang, Y.~Dong, T.~Pang, Z.~Xiao, H.~Su, and J.~Zhu, ``Controllable evaluation and generation of physical adversarial patch on face recognition,'' \emph{arXiv preprint arXiv:2203.04623}, 2022.

\bibitem{PadvFace}
X.~Zheng, Y.~Fan, B.~Wu, Y.~Zhang, J.~Wang, and S.~Pan, ``Robust physical-world attacks on face recognition,'' \emph{Pattern Recognition}, vol. 133, p. 109009, 2023.

\bibitem{wei2022simultaneously}
X.~Wei, Y.~Guo, J.~Yu, and B.~Zhang, ``Simultaneously optimizing perturbations and positions for black-box adversarial patch attacks,'' \emph{IEEE transactions on pattern analysis and machine intelligence}, 2022.

\bibitem{DOPatch}
X.~Wei, S.~Ruan, Y.~Dong, and H.~Su, ``Distributional modeling for location-aware adversarial patches,'' \emph{arXiv preprint arXiv:2306.16131}, 2023.

\bibitem{SASMask}
H.~Gong, M.~Dong, S.~Ma, S.~Camtepe, S.~Nepal, and C.~Xu, ``Stealthy physical masked face recognition attack via adversarial style optimization,'' \emph{IEEE Transactions on Multimedia}, 2023.

\bibitem{RSTAM}
X.~Liu, F.~Shen, J.~Zhao, and C.~Nie, ``Rstam: An effective black-box impersonation attack on face recognition using a mobile and compact printer,'' \emph{arXiv preprint arXiv:2206.12590}, 2022.

\bibitem{hwang2023adversarial}
R.-H. Hwang, J.-Y. Lin, S.-Y. Hsieh, H.-Y. Lin, and C.-L. Lin, ``Adversarial patch attacks on deep-learning-based face recognition systems using generative adversarial networks,'' \emph{Sensors}, vol.~23, no.~2, p. 853, 2023.

\bibitem{EAP}
X.~Liu, F.~Shen, J.~Zhao, and C.~Nie, ``Eap: An effective black-box impersonation adversarial patch attack method on face recognition in the physical world,'' \emph{Neurocomputing}, vol. 580, p. 127517, 2024.

\bibitem{AGNs}
M.~Sharif, S.~Bhagavatula, L.~Bauer, and M.~K. Reiter, ``A general framework for adversarial examples with objectives,'' \emph{ACM Transactions on Privacy and Security (TOPS)}, vol.~22, no.~3, pp. 1--30, 2019.

\bibitem{glass-noise}
I.~Singh, T.~Araki, and K.~Kakizaki, ``Powerful physical adversarial examples against practical face recognition systems,'' in \emph{Proceedings of the IEEE/CVF Winter Conference on Applications of Computer Vision}, 2022, pp. 301--310.

\bibitem{cohen2023accessorize}
A.~Cohen and M.~Sharif, ``Accessorize in the dark: A security analysis of near-infrared face recognition,'' in \emph{European Symposium on Research in Computer Security}.\hskip 1em plus 0.5em minus 0.4em\relax Springer, 2023, pp. 43--61.

\bibitem{singh2021brightness}
I.~Singh, S.~Momiyama, K.~Kakizaki, and T.~Araki, ``On brightness agnostic adversarial examples against face recognition systems,'' in \emph{2021 International Conference of the Biometrics Special Interest Group (BIOSIG)}.\hskip 1em plus 0.5em minus 0.4em\relax IEEE, 2021, pp. 1--5.

\bibitem{CC2023}
{CAI Chuxin }, { WANG Yufei }, { ZHANG Liepiao }, { ZHUO Sichao }, { ZHANG Juanmiao }, and { HU Yongjian}, ``Adversarial attacks on face recognition system in physical domain,'' \emph{Journal of Cyber Security}, vol.~8, no.~2, pp. 127--137, 2023.

\bibitem{makeup2}
C.-S. Lin, C.-Y. Hsu, P.-Y. Chen, and C.-M. Yu, ``Real-world adversarial examples via makeup,'' in \emph{ICASSP 2022-2022 IEEE International Conference on Acoustics, Speech and Signal Processing (ICASSP)}.\hskip 1em plus 0.5em minus 0.4em\relax IEEE, 2022, pp. 2854--2858.

\bibitem{AdvEye}
J.~Pi, J.~Zeng, Q.~Lu, N.~Jiang, H.~Wu, L.~Zeng, and Z.~Wu, ``Adv-eye: A transfer-based natural eye shadow attack on face recognition,'' \emph{IEEE Access}, 2023.

\bibitem{ImU}
S.~An, Y.~Yao, Q.~Xu, S.~Ma, G.~Tao, S.~Cheng, K.~Zhang, Y.~Liu, G.~Shen, I.~Kelk \emph{et~al.}, ``Imu: Physical impersonating attack for face recognition system with natural style changes,'' in \emph{2023 IEEE Symposium on Security and Privacy (SP)}.\hskip 1em plus 0.5em minus 0.4em\relax IEEE, 2023, pp. 899--916.

\bibitem{Face++}
\BIBentryALTinterwordspacing
{Face++}. [Online]. Available: \url{https://www.faceplusplus.com}
\BIBentrySTDinterwordspacing

\bibitem{Aliyun}
\BIBentryALTinterwordspacing
{Aliyun}. [Online]. Available: \url{https://vision.aliyun.com/facebody}
\BIBentrySTDinterwordspacing

\bibitem{Microsoft}
\BIBentryALTinterwordspacing
{Microsoft}. [Online]. Available: \url{https://azure.microsoft.com/en-us/services/cognitive-services/face}
\BIBentrySTDinterwordspacing

\bibitem{Clarifai}
\BIBentryALTinterwordspacing
{Clarifai}. [Online]. Available: \url{https://www.clarifai.com/}
\BIBentrySTDinterwordspacing

\bibitem{Baidu}
\BIBentryALTinterwordspacing
Baidu. [Online]. Available: \url{https://ai.baidu.com/tech/face}
\BIBentrySTDinterwordspacing

\bibitem{Tencent}
\BIBentryALTinterwordspacing
{Tencent}. [Online]. Available: \url{https://cloud.tencent.com/product/facerecognition}
\BIBentrySTDinterwordspacing

\bibitem{mask}
A.~Zolfi, S.~Avidan, Y.~Elovici, and A.~Shabtai, ``Adversarial mask: Real-world universal adversarial attack on face recognition models,'' in \emph{Joint European Conference on Machine Learning and Knowledge Discovery in Databases}.\hskip 1em plus 0.5em minus 0.4em\relax Springer, 2022, pp. 304--320.

\bibitem{checkerboard}
C.~Zhou, H.~Jing, X.~He, L.~Wang, K.~Chen, and D.~Ma, ``Disappeared face: a physical adversarial attack method on black-box face detection models,'' in \emph{Information and Communications Security: 23rd International Conference, ICICS 2021, Chongqing, China, November 19-21, 2021, Proceedings, Part I 23}.\hskip 1em plus 0.5em minus 0.4em\relax Springer, 2021, pp. 119--135.

\bibitem{AT3D}
X.~Yang, C.~Liu, L.~Xu, Y.~Wang, Y.~Dong, N.~Chen, H.~Su, and J.~Zhu, ``Towards effective adversarial textured 3d meshes on physical face recognition,'' in \emph{Proceedings of the IEEE/CVF Conference on Computer Vision and Pattern Recognition}, 2023, pp. 4119--4128.

\bibitem{Projection}
D.-L. Nguyen, S.~S. Arora, Y.~Wu, and H.~Yang, ``Adversarial light projection attacks on face recognition systems: A feasibility study,'' in \emph{Proceedings of the IEEE/CVF conference on computer vision and pattern recognition workshops}, 2020, pp. 814--815.

\bibitem{Optical}
Y.~Li, Y.~Li, X.~Dai, S.~Guo, and B.~Xiao, ``Physical-world optical adversarial attacks on 3d face recognition,'' in \emph{Proceedings of the IEEE/CVF Conference on Computer Vision and Pattern Recognition}, 2023, pp. 24\,699--24\,708.

\bibitem{LFW}
G.~B. Huang, M.~Mattar, T.~Berg, and E.~Learned-Miller, ``Labeled faces in the wild: A database forstudying face recognition in unconstrained environments,'' in \emph{Workshop on faces in'Real-Life'Images: detection, alignment, and recognition}, 2008.

\bibitem{PubFig}
N.~Kumar, A.~C. Berg, P.~N. Belhumeur, and S.~K. Nayar, ``Attribute and simile classifiers for face verification,'' in \emph{2009 IEEE 12th international conference on computer vision}.\hskip 1em plus 0.5em minus 0.4em\relax IEEE, 2009, pp. 365--372.

\bibitem{YMU}
A.~Dantcheva, C.~Chen, and A.~Ross, ``Can facial cosmetics affect the matching accuracy of face recognition systems?'' in \emph{2012 IEEE Fifth international conference on biometrics: theory, applications and systems (BTAS)}.\hskip 1em plus 0.5em minus 0.4em\relax IEEE, 2012, pp. 391--398.

\bibitem{li2013casia}
S.~Li, D.~Yi, Z.~Lei, and S.~Liao, ``The casia nir-vis 2.0 face database,'' in \emph{Proceedings of the IEEE conference on computer vision and pattern recognition workshops}, 2013, pp. 348--353.

\bibitem{CASIA-WebFace}
D.~Yi, Z.~Lei, S.~Liao, and S.~Z. Li, ``Learning face representation from scratch,'' \emph{arXiv preprint arXiv:1411.7923}, 2014.

\bibitem{CelebA}
Z.~Liu, P.~Luo, X.~Wang, and X.~Tang, ``Deep learning face attributes in the wild,'' in \emph{Proceedings of the IEEE international conference on computer vision}, 2015, pp. 3730--3738.

\bibitem{WiderFace}
S.~Yang, P.~Luo, C.-C. Loy, and X.~Tang, ``Wider face: A face detection benchmark,'' in \emph{Proceedings of the IEEE conference on computer vision and pattern recognition}, 2016, pp. 5525--5533.

\bibitem{CFP}
S.~Sengupta, J.-C. Chen, C.~Castillo, V.~M. Patel, R.~Chellappa, and D.~W. Jacobs, ``Frontal to profile face verification in the wild,'' in \emph{2016 IEEE winter conference on applications of computer vision (WACV)}.\hskip 1em plus 0.5em minus 0.4em\relax IEEE, 2016, pp. 1--9.

\bibitem{CASIA-FaceV5}
{CAS Institute of Automation}, ``Casia-facev5,'' \url{http://biometrics.idealtest.org}, 2009, last accessed on 2024-08-5.

\bibitem{VGGFace2}
Q.~Cao, L.~Shen, W.~Xie, O.~M. Parkhi, and A.~Zisserman, ``Vggface2: A dataset for recognising faces across pose and age,'' in \emph{2018 13th IEEE international conference on automatic face \& gesture recognition (FG 2018)}.\hskip 1em plus 0.5em minus 0.4em\relax IEEE, 2018, pp. 67--74.

\bibitem{AgeDB}
S.~Moschoglou, A.~Papaioannou, C.~Sagonas, J.~Deng, I.~Kotsia, and S.~Zafeiriou, ``Agedb: the first manually collected, in-the-wild age database,'' in \emph{proceedings of the IEEE conference on computer vision and pattern recognition workshops}, 2017, pp. 51--59.

\bibitem{MS-Celeb}
Y.~Guo, L.~Zhang, Y.~Hu, X.~He, and J.~Gao, ``Ms-celeb-1m: A dataset and benchmark for large-scale face recognition,'' in \emph{Computer Vision--ECCV 2016: 14th European Conference, Amsterdam, The Netherlands, October 11-14, 2016, Proceedings, Part III 14}.\hskip 1em plus 0.5em minus 0.4em\relax Springer, 2016, pp. 87--102.

\bibitem{CelebA-HQ}
T.~Karras, T.~Aila, S.~Laine, and J.~Lehtinen, ``Progressive growing of gans for improved quality, stability, and variation,'' \emph{arXiv preprint arXiv:1710.10196}, 2017.

\bibitem{MT}
T.~Li, R.~Qian, C.~Dong, S.~Liu, Q.~Yan, W.~Zhu, and L.~Lin, ``Beautygan: Instance-level facial makeup transfer with deep generative adversarial network,'' in \emph{Proceedings of the 26th ACM international conference on Multimedia}, 2018, pp. 645--653.

\bibitem{K-Face}
Y.~Choi, H.~Park, G.~P. Nam, H.~Kim, H.~Choi, J.~Cho, and I.-J. Kim, ``K-face: A large-scale kist face database in consideration with unconstrained environments,'' \emph{arXiv preprint arXiv:2103.02211}, 2021.

\bibitem{LADN}
Q.~Gu, G.~Wang, M.~T. Chiu, Y.-W. Tai, and C.-K. Tang, ``Ladn: Local adversarial disentangling network for facial makeup and de-makeup,'' in \emph{Proceedings of the IEEE/CVF International conference on computer vision}, 2019, pp. 10\,481--10\,490.

\bibitem{Bosphorus}
A.~Savran, N.~Aly{\"u}z, H.~Dibeklio{\u{g}}lu, O.~{\c{C}}eliktutan, B.~G{\"o}kberk, B.~Sankur, and L.~Akarun, ``Bosphorus database for 3d face analysis,'' in \emph{Biometrics and Identity Management: First European Workshop, BIOID 2008, Roskilde, Denmark, May 7-9, 2008. Revised Selected Papers 1}.\hskip 1em plus 0.5em minus 0.4em\relax Springer, 2008, pp. 47--56.

\bibitem{Eurecom}
R.~Min, N.~Kose, and J.-L. Dugelay, ``Kinectfacedb: A kinect database for face recognition,'' \emph{IEEE Transactions on Systems, Man, and Cybernetics: Systems}, vol.~44, no.~11, pp. 1534--1548, 2014.

\bibitem{SIAT-3DFE}
Y.~Ye, Z.~Song, J.~Guo, and Y.~Qiao, ``Siat-3dfe: a high-resolution 3d facial expression dataset,'' \emph{IEEE Access}, vol.~8, pp. 48\,205--48\,211, 2020.

\bibitem{yamada2013privacy}
T.~Yamada, S.~Gohshi, and I.~Echizen, ``Privacy visor: Method based on light absorbing and reflecting properties for preventing face image detection,'' in \emph{2013 IEEE International Conference on Systems, Man, and Cybernetics}.\hskip 1em plus 0.5em minus 0.4em\relax IEEE, 2013, pp. 1572--1577.

\bibitem{FacePET}
A.~J. Perez, S.~Zeadally, L.~Y. Matos~Garcia, J.~A. Mouloud, and S.~Griffith, ``Facepet: Enhancing bystanders’ facial privacy with smart wearables/internet of things,'' \emph{Electronics}, vol.~7, no.~12, p. 379, 2018.

\bibitem{tvloss}
L.~I. Rudin, S.~Osher, and E.~Fatemi, ``Nonlinear total variation based noise removal algorithms,'' \emph{Physica D: nonlinear phenomena}, vol.~60, no. 1-4, pp. 259--268, 1992.

\bibitem{EOT}
A.~Athalye, L.~Engstrom, A.~Ilyas, and K.~Kwok, ``Synthesizing robust adversarial examples,'' in \emph{International conference on machine learning}.\hskip 1em plus 0.5em minus 0.4em\relax PMLR, 2018, pp. 284--293.

\bibitem{3DMM}
A.~T. Tran, T.~Hassner, I.~Masi, and G.~Medioni, ``Regressing robust and discriminative 3d morphable models with a very deep neural network,'' in \emph{2017 IEEE Conference on Computer Vision and Pattern Recognition (CVPR)}.\hskip 1em plus 0.5em minus 0.4em\relax IEEE, 2017, pp. 1493--1502.

\bibitem{pytorch3d}
J.~Johnson, N.~Ravi, J.~Reizenstein, D.~Novotny, S.~Tulsiani, C.~Lassner, and S.~Branson, ``Accelerating 3d deep learning with pytorch3d,'' in \emph{SIGGRAPH Asia 2020 courses}, 2020, pp. 1--1.

\bibitem{xie2019improving}
C.~Xie, Z.~Zhang, Y.~Zhou, S.~Bai, J.~Wang, Z.~Ren, and A.~L. Yuille, ``Improving transferability of adversarial examples with input diversity,'' in \emph{Proceedings of the IEEE/CVF conference on computer vision and pattern recognition}, 2019, pp. 2730--2739.

\bibitem{tramer2017ensemble}
F.~Tramèr, A.~Kurakin, N.~Papernot, I.~Goodfellow, D.~Boneh, and P.~McDaniel, ``Ensemble adversarial training: Attacks and defenses,'' in \emph{International Conference on Learning Representations}, 2018.

\bibitem{liu2016delving}
Y.~Liu, X.~Chen, C.~Liu, and D.~Song, ``Delving into transferable adversarial examples and black-box attacks,'' in \emph{International Conference on Learning Representations}, 2017.

\bibitem{WGAN-GP}
I.~Gulrajani, F.~Ahmed, M.~Arjovsky, V.~Dumoulin, and A.~C. Courville, ``Improved training of wasserstein gans,'' \emph{Advances in neural information processing systems}, vol.~30, 2017.

\bibitem{G3D}
Y.~Shi, D.~Aggarwal, and A.~K. Jain, ``Lifting 2d stylegan for 3d-aware face generation,'' in \emph{Proceedings of the IEEE/CVF conference on computer vision and pattern recognition}, 2021, pp. 6258--6266.

\bibitem{Cycle-GAN}
J.-Y. Zhu, T.~Park, P.~Isola, and A.~A. Efros, ``Unpaired image-to-image translation using cycle-consistent adversarial networks,'' in \emph{Proceedings of the IEEE international conference on computer vision}, 2017, pp. 2223--2232.

\bibitem{LPIPS}
R.~Zhang, P.~Isola, A.~A. Efros, E.~Shechtman, and O.~Wang, ``The unreasonable effectiveness of deep features as a perceptual metric,'' in \emph{Proceedings of the IEEE conference on computer vision and pattern recognition}, 2018, pp. 586--595.

\bibitem{SSIM}
Z.~Wang, A.~C. Bovik, H.~R. Sheikh, and E.~P. Simoncelli, ``Image quality assessment: from error visibility to structural similarity,'' \emph{IEEE transactions on image processing}, vol.~13, no.~4, pp. 600--612, 2004.

\bibitem{AlexNet}
A.~Krizhevsky, I.~Sutskever, and G.~E. Hinton, ``Imagenet classification with deep convolutional neural networks,'' \emph{Communications of the ACM}, vol.~60, no.~6, pp. 84--90, 2017.

\bibitem{Meta-learning}
R.~Shao, X.~Lan, and P.~C. Yuen, ``Regularized fine-grained meta face anti-spoofing,'' in \emph{Proceedings of the AAAI conference on artificial intelligence}, vol.~34, no.~07, 2020, pp. 11\,974--11\,981.

\bibitem{TaintRadar}
F.~Li, X.~Liu, X.~Zhang, Q.~Li, K.~Sun, and K.~Li, ``Detecting localized adversarial examples: A generic approach using critical region analysis,'' in \emph{IEEE INFOCOM 2021-IEEE Conference on Computer Communications}.\hskip 1em plus 0.5em minus 0.4em\relax IEEE, 2021, pp. 1--10.

\bibitem{SentiNet}
E.~Chou, F.~Tramer, and G.~Pellegrino, ``Sentinet: Detecting localized universal attacks against deep learning systems,'' in \emph{2020 IEEE Security and Privacy Workshops (SPW)}.\hskip 1em plus 0.5em minus 0.4em\relax IEEE, 2020, pp. 48--54.

\bibitem{xie2023random}
J.~Xie, Y.~Luo, and J.~Lu, ``A random-patch based defense strategy against physical attacks for face recognition systems,'' \emph{arXiv preprint arXiv:2304.07822}, 2023.

\bibitem{EAD}
L.~Wu, X.~Yang, Y.~Dong, L.~XIE, H.~Su, and J.~Zhu, ``Embodied active defense: Leveraging recurrent feedback to counter adversarial patches,'' in \emph{International Conference on Learning Representations}, 2024.

\bibitem{peng2024detection}
F.~Peng, L.~Qin, M.~Long, and J.~Li, ``Detection of adversarial facial accessory presentation attacks using local face differential,'' \emph{ACM Transactions on Multimedia Computing, Communications and Applications}, vol.~20, no.~7, pp. 1--28, 2024.

\bibitem{RADAP}
X.~Liu, F.~Shen, J.~Zhao, and C.~Nie, ``Radap: A robust and adaptive defense against diverse adversarial patches on face recognition,'' \emph{arXiv preprint arXiv:2311.17339}, 2023.

\bibitem{AmI}
G.~Tao, S.~Ma, Y.~Liu, and X.~Zhang, ``Attacks meet interpretability: Attribute-steered detection of adversarial samples,'' \emph{Advances in neural information processing systems}, vol.~31, 2018.

\bibitem{UniFAD}
D.~Deb, X.~Liu, and A.~K. Jain, ``Unified detection of digital and physical face attacks,'' in \emph{2023 IEEE 17th International Conference on Automatic Face and Gesture Recognition (FG)}, 2023, pp. 1--8.

\bibitem{DIFFender}
C.~Kang, Y.~Dong, Z.~Wang, S.~Ruan, H.~Su, and X.~Wei, ``Diffender: Diffusion-based adversarial defense against patch attacks in the physical world,'' in \emph{International Conference on Learning Representations}, 2024.

\bibitem{PIN}
M.~Ren, Y.~Zhu, Y.~Wang, and Z.~Sun, ``Perturbation inactivation based adversarial defense for face recognition,'' \emph{IEEE Transactions on Information Forensics and Security}, vol.~17, pp. 2947--2962, 2022.

\bibitem{Jujutsu}
Z.~Chen, P.~Dash, and K.~Pattabiraman, ``Jujutsu: A two-stage defense against adversarial patch attacks on deep neural networks,'' in \emph{Proceedings of the 2023 ACM Asia Conference on Computer and Communications Security}, 2023, pp. 689--703.

\bibitem{DOA}
T.~Wu, L.~Tong, and Y.~Vorobeychik, ``Defending against physically realizable attacks on image classification,'' in \emph{International Conference on Learning Representations}, 2020.

\bibitem{arvinte2020robust}
M.~Arvinte, A.~H. Tewfik, and S.~Vishwanath, ``Robust face verification via disentangled representations,'' \emph{arXiv preprint arXiv:2006.03638}, 2020.

\bibitem{ren2024artificial}
M.~Ren, Y.~Wang, Y.~Zhu, Y.~Huang, Z.~Sun, Q.~Li, and T.~Tan, ``Artificial immune system of secure face recognition against adversarial attacks,'' \emph{International Journal of Computer Vision}, pp. 1--23, 2024.

\bibitem{chiang2020certified}
P.~yeh Chiang, R.~Ni, A.~Abdelkader, C.~Zhu, C.~Studor, and T.~Goldstein, ``Certified defenses for adversarial patches,'' in \emph{International Conference on Learning Representations}, 2020.

\bibitem{zhang2020clipped}
Z.~Zhang, B.~Yuan, M.~McCoyd, and D.~Wagner, ``Clipped bagnet: Defending against sticker attacks with clipped bag-of-features,'' in \emph{2020 IEEE Security and Privacy Workshops (SPW)}.\hskip 1em plus 0.5em minus 0.4em\relax IEEE, 2020, pp. 55--61.

\bibitem{xiang2021detectorguard}
C.~Xiang and P.~Mittal, ``Detectorguard: Provably securing object detectors against localized patch hiding attacks,'' in \emph{Proceedings of the 2021 ACM SIGSAC Conference on Computer and Communications Security}, 2021, pp. 3177--3196.

\bibitem{chen2022towards}
Z.~Chen, B.~Li, J.~Xu, S.~Wu, S.~Ding, and W.~Zhang, ``Towards practical certifiable patch defense with vision transformer,'' in \emph{Proceedings of the IEEE/CVF Conference on Computer Vision and Pattern Recognition}, 2022, pp. 15\,148--15\,158.

\bibitem{xiang2022patchcleanser}
C.~Xiang, S.~Mahloujifar, and P.~Mittal, ``$\{$PatchCleanser$\}$: Certifiably robust defense against adversarial patches for any image classifier,'' in \emph{31st USENIX Security Symposium (USENIX Security 22)}, 2022, pp. 2065--2082.

\bibitem{gowal2018effectiveness}
S.~Gowal, K.~Dvijotham, R.~Stanforth, R.~Bunel, C.~Qin, J.~Uesato, R.~Arandjelovic, T.~Mann, and P.~Kohli, ``On the effectiveness of interval bound propagation for training verifiably robust models,'' \emph{arXiv preprint arXiv:1810.12715}, 2018.

\bibitem{mirman2018differentiable}
M.~Mirman, T.~Gehr, and M.~Vechev, ``Differentiable abstract interpretation for provably robust neural networks,'' in \emph{International Conference on Machine Learning}.\hskip 1em plus 0.5em minus 0.4em\relax PMLR, 2018, pp. 3578--3586.

\bibitem{10445007}
J.~Zhang, J.~Huang, S.~Jin, and S.~Lu, ``Vision-language models for vision tasks: A survey,'' \emph{IEEE Transactions on Pattern Analysis and Machine Intelligence}, vol.~46, no.~8, pp. 5625--5644, 2024.

\bibitem{liu2024adv}
D.~Liu, X.~Wang, C.~Peng, N.~Wang, R.~Hu, and X.~Gao, ``Adv-diffusion: imperceptible adversarial face identity attack via latent diffusion model,'' in \emph{Proceedings of the AAAI Conference on Artificial Intelligence}, vol.~38, no.~4, 2024, pp. 3585--3593.

\bibitem{hu2024towards}
C.~Hu, Y.~Li, Z.~Feng, and X.~Wu, ``Towards transferable attack via adversarial diffusion in face recognition,'' \emph{IEEE Transactions on Information Forensics and Security}, 2024.

\bibitem{bergstra2012random}
J.~Bergstra and Y.~Bengio, ``Random search for hyper-parameter optimization.'' \emph{Journal of machine learning research}, vol.~13, no.~2, 2012.

\bibitem{snoek2012practical}
J.~Snoek, H.~Larochelle, and R.~P. Adams, ``Practical bayesian optimization of machine learning algorithms,'' \emph{Advances in neural information processing systems}, vol.~25, 2012.

\bibitem{akiba2019optuna}
T.~Akiba, S.~Sano, T.~Yanase, T.~Ohta, and M.~Koyama, ``Optuna: A next-generation hyperparameter optimization framework,'' in \emph{Proceedings of the 25th ACM SIGKDD international conference on knowledge discovery \& data mining}, 2019, pp. 2623--2631.

\end{thebibliography}

% \newpage
% \section{Biography Section}
% \begin{IEEEbiography}[{\includegraphics[width=1in,height=1.25in,clip,keepaspectratio]{fig1}}]{Michael Shell}
% Use $\backslash${\tt{begin{IEEEbiography}}} and then for the 1st argument use $\backslash${\tt{includegraphics}} to declare and link the author photo.
% Use the author name as the 3rd argument followed by the biography text.
% \end{IEEEbiography}

% \bf{If you will not include a photo:}
% \begin{IEEEbiographynophoto}{John Doe}
% Use $\backslash${\tt{begin{IEEEbiographynophoto}}} and the author name as the argument followed by the biography text.
% \end{IEEEbiographynophoto}

\vfill

\end{document}